\documentclass[conference]{IEEEtran}

\usepackage[normalem]{ulem}
\usepackage{epsfig}
\usepackage{graphicx}
\usepackage{balance}
\usepackage{comment}
\usepackage{textcomp}
\usepackage{listings}
\usepackage{color,soul}
\usepackage{colortbl}
\usepackage{leading}
\usepackage{moreverb}
\usepackage{listings}
\usepackage{framed}
\usepackage{multirow}
\usepackage[hyphens]{url}
\usepackage{wrapfig}

\usepackage{tikz}

\makeatletter
\newif\if@restonecol
\makeatother

\usepackage[boxed]{algorithm2e}
\definecolor{lightgray}{gray}{0.9}
\definecolor{lightblue}{rgb}{0.9,0.9,1}
\definecolor{red}{rgb}{1,0,0}

\newcommand{\etc}{\emph{etc.}\xspace}
\newcommand{\ie}{\emph{i.e.,}\xspace}
\newcommand{\eg}{\emph{e.g.,}\xspace}

\usepackage{subcaption}
\usepackage{tcolorbox}
\usepackage{paralist, tabularx, booktabs}
\usepackage{makecell}

\newcounter{fnum}
\newcommand{\nextfnum}{\refstepcounter{fnum}\arabic{fnum}}

\begin{document}

\title{SyzRetrospector: A Large-Scale Retrospective Study of Syzbot}

\author{
\IEEEauthorblockN{Joseph Bursey}
\IEEEauthorblockA{University of California, Irvine\\
jbursey@uci.edu}
\and
\IEEEauthorblockN{Ardalan Amiri Sani}
\IEEEauthorblockA{University of California, Irvine\\
ardalan@uci.edu}
\and
\IEEEauthorblockN{Zhiyun Qian}
\IEEEauthorblockA{University of California, Riverside\\
zhiyunq@cs.ucr.edu}
}

\maketitle

\pagestyle{plain}
 
\begin{abstract}
Over the past $6$ years, Syzbot has fuzzed the Linux kernel day and night to report over $5570$ bugs, of which $4604$ have been patched~\cite{syzbot}.
While this is impressive, we have found the average time to find a bug is over $405$ days.
Moreover, we have found that current metrics commonly used, such as time-to-find and number of bugs found, are inaccurate in evaluating Syzbot since bugs often spend the majority of their lives hidden from the fuzzer.
In this paper, we set out to better understand and quantify Syzbot's performance and improvement in finding bugs.
Our tool, SyzRetrospector, takes a different approach to evaluating Syzbot by finding the earliest that Syzbot was capable of finding a bug, and why that bug was revealed.
We use SyzRetrospector on a large scale to analyze $559$ bugs and find that bugs are hidden for an average of $331.17$ days before Syzbot is even able to find them.
We further present findings on the behaviors of revealing factors, how some bugs are harder to reveal than others, the trends in delays over the past $6$ years, and how bug location relates to delays.
We also provide key takeaways for improving Syzbot's delays.
\end{abstract}

\section{Introduction}
\label{sec:introduction}

Over the past $6$ years, Syzbot, one of the largest continuous fuzzing projects, has managed to find over $5570$ bugs~\cite{syzbot} in the Linux kernel.
Yet despite this number, the question ``How well has Syzbot performed?'' cannot be precisely answered.
Syzbot has found $2.58$ bugs per day over its lifetime, but we do not know how many bugs were missed.
Is Syzbot finding bugs faster than they are being introduced?
What is limiting it from finding more bugs?

One way to evaluate the fuzzer takes advantage of its continuous nature.
We can find a bug's overall finding delay by counting the days from when it was introduced to when it was found.
A bug's delay will tell us how well Syzbot performed in finding that bug, where a shorter delay is better.
This metric is also detached from the number of bugs Syzbot has or has not found since it is based on the bug's lifetime.
Bugs that are not found yet or are too hard to find would only lengthen the average delay, so the measured delay gives a lower bound.
However, in a small pilot study (\S\ref{subsec:pilot_study}), we find that this overall delay is inaccurate.
We find that bugs are not always findable by Syzbot, sometimes due to the fuzzer's capability or code in the kernel hiding the bug.
Indeed, $68.16\%$ of bugs found by Syzbot have spent some amount of their lifetimes hidden from the fuzzer.

Based on this obervation, we divide the overall delay into two parts: one where the bug is hidden from the fuzzer, and another where the bug is revealed, but not yet found.
We call these delays $D_1$ and $D_2$ respectively, and use them as key metrics in our study.
These delays have the advantage of being rooted in the context of a bug's lifetime, while still being true to what the fuzzer is capable of.

We present SyzRetrospector, a tool purpose-built to identify $D_1$ and $D_2$ by finding the dividing line between them.
SyzRetrospector is capable of going back in time to faithfully recreate Syzbot's fuzzing environment at a given point in time in order to find the exact day and reason Syzbot was first able to find a bug.
The resulting reason, which we call the \textit{revealing factor}, is the dividing line between the two delays.
Over a $3.5$ month period, we used SyzRetrospector on a large scale to find the revealing factors of $559$ bugs\footnote{We will open source SyzRetrospector and all our study results}.
With a solid understanding of how long bugs are hidden for and how long until they are found afterwards, we present our evaluation of Syzbot's performance over the past $6$ years.

We identify $5$ different revealing factors, each with different behaviors, and highlight that some may be hard to improve such that the bug finding delay decreases.
We present a breakdown of the bug finding delay as $D_1$ and $D_2$, and show how they can be used to evaluate the fuzzer over time.
We then provide strategies for improving $D_1$ and $D_2$ such as using CVEs to find gaps in Syzbot's syscall description set.

In summary, we make the following contributions.
\begin{compactitem}
    \item
    We carry out the first large-scale, dynamic measurement study of Syzbot's performance.
    \item
    We introduce SyzRetrospector, a tool capable of analyzing the lifetime of a bug and its findability with respect to Syzbot.
    \item
    We present several important findings about the bug finding delay performance of Syzbot.
    \item
    We suggest key takeaways that provide a structure for improving Syzbot.
\end{compactitem}

\section{Background}

\subsection{Fuzzing}
Fuzzing is a debugging process by which randomized inputs are passed to a target program and run in the hopes of finding bugs, which manifest as crashes.
Fuzzing engines, or fuzzers, often gain additional feedback from the target program such as code coverage, which can be used for future input generation or mutation.
Plenty of research and development has gone into fuzzers to make them more efficient at bug finding, which has earned them a prominent role in debugging.

\subsection{Continuous Fuzzing and Syzbot}
Continuous fuzzers can run non-stop alongside development of long-standing projects like Linux, finding bugs as they are introduced.
Such fuzzers can amass a large corpus of test cases and maintain a complex fuzzer state across years of testing.
Continuous fuzzing can give a distict advantage when more time is available.

Syzbot, Google's coverage-guided continuous fuzzer, began fuzzing the Linux kernel in 2017, and continues today with 24 instances called \textit{managers}, which are instances of Syzkaller.
To be clear, Syzkaller interfaces with and fuzzes the kernel, while Syzbot is the apparatus that allows Syzkaller to fuzz continuously and communicate with other managers.
Syzbot oversees each instance of Syzkaller, sharing test cases between them and periodically updating both Linux and Syzkaller.
This way, Syzbot is always fuzzing relevant kernel versions and has the most up to date version of Syzkaller to fuzz with.

Each of the 24 managers is configured differently to cover different modules, sanitizers, architectures, or repositories.
For example, the manager \textit{ci-upstream-kasan-gce} fuzzes the upstream Linux respository with KASAN active~\cite{syzbot}.
KASAN is one of the several sanitizers used by Syzbot, which we will provide background for in \S\ref{subsec:sanitizers}.
It should be noted that most managers have as many modules active as possible, barring conflicts with other modules, and many use KASAN, even if it is not stated.

A Syzkaller instance begins by generating a number of test cases or inputs from syscall descriptions which are manually written and built into the fuzzer.
Syzkaller runs these inputs on the target kernel and adds to the corpus those that provide new coverage or generate a crash.
It then selects inputs from the corpus to be mutated and run again, and the cycle repeats.
With this strategy, Syzbot has solidified its position in the Linux development cycle as a key tool for bug finding.

\subsection{Sanitizers}
\label{subsec:sanitizers}
Sanitizers are bug finding tools that are built into the kernel.
They instrument the kernel code and allow the fuzzer to find specific types of bugs, such as memory bugs.
They are particularly useful to fuzzers because they generate crashes from bugs that would normally go unnoticed.
For instance, a pointer use-after-free would not normally generate a crash, thus hiding it from a fuzzer.
Instead, KASAN (Kernel Adress SANitizer)~\cite{kasan2023}, identifies the use-after-free and generates a crash, which the fuzzer reports.
Syzbot makes use of many of these sanitizers such as KASAN, UBSAN (Undefined Behavior SANitizer)~\cite{ubsan2023}, KMSAN (Kernel Memory SANitizer)~\cite{kmsan2023}, and KCSAN (Kernel Concurrency SANitizer)~\cite{kcsan2023}.
These sanitizers trigger on undefined behavior, uninitialized values, and race conditions respectively.
In order to find memory leaks, Syzbot uses the Kernel Memory Leak Detector, known as KMEMLEAK~\cite{kmemleak2023}.
These sanitizers are used to greatly increase the types of bugs Syzbot can find.

\section{Motivation and Definitions}
\label{sec:motivation}

Here we present our motivation through a pilot study (\S\ref{subsec:pilot_study}), in which we identify the different reasons a bug could be hidden, as well as the dates that describe the bug's life cycle.
We then provide examples which demonstrate how revealing factors reveal bugs (\S\ref{subsec:revealing_factors}).

\begin{figure}
    \centering
    \includegraphics[width=\columnwidth]{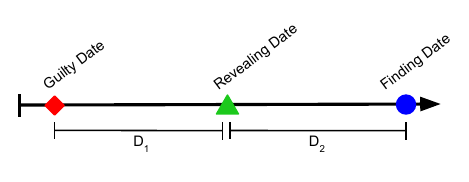}
    \vspace{-0.3in}
    \caption{The lifetime of a bug.}
    \vspace{-0.2in}
    \label{fig:lifetime}
\end{figure}

\subsection{Pilot Study}
\label{subsec:pilot_study}

In a manual pilot study of $20$ recent bugs, we identified several reasons a bug could be hidden from Syzbot, as well as three key dates that describe the life cycle of the bugs (Figure~\ref{fig:lifetime}).
First, every bug has a commit in which it was introduced to the kernel - its \textit{Guilty Commit}.
In some cases, this commit is a merge commit that pulls code from another repository.
Our study will focus only on bugs found in the upstream (i.e., mainline) Linux repository as we will explain in \S\ref{subsec:repo_choice}.
Regardless of whether it was introduced by merge or not, we mark the day a bug is introduced to upstream as the \textit{Guilty Date}.

At the guilty date, the bug may or may not be findable by the fuzzer.
In fact, our study shows that $9/20$ were not findable yet.
The hidden bugs lie in wait until they are revealed by a revealing factor, which will be explained in great detail in \S\ref{subsec:revealing_factors}.
They range from syscall description updates and other Syzkaller commits, to changes in kernel code or blocking bugs.

We can enumerate each of these \textit{Revealing Factors}, and precisely identify \textit{Revealing Commits}.
A bug's \textit{Revealing Commit} is the commit that enables the fuzzer to find that bug.
Based on the revealing factors described below, this commit can be in either Linux or Syzkaller.
In our pilot study we found $3$ bugs revealed by changes in kernel code, $2$ bugs that were hidden behind blocking bugs, $3$ bugs that were unfindable until a syscall description update, and $1$ that was found due to a general Syzkaller update.
We also add one more revealing factor not seen in our pilot study denoted by a sanitizer commit.
The day a bug is revealed marks its \textit{Revealing Date}, after which the fuzzer can find it.

Lastly, the \textit{Finding Date} is the date when a bug is found.
On the finding date, Syzbot was fuzzing a particular Linux commit, which we will call the \textit{Finding Commit}.
Not only does this commit mark the end of the bug's finding delay, but it can also act as an anchor for our study since we know the bug reproduces in this commit.

Using these dates, we get two periods of time in a bug's life cycle.
We define $D_1$ as the period of time where the bug is hidden or unfindable (\textit{Revealing Date} - \textit{Guilty Date}).
As a part of our design (\S\ref{sec:design}), and in order to faithfully recreate Syzbot's environment through SyzRetrospector, $D_1$ will be truncated by the date Syzbot first began fuzzing.
Syzbot was unable to fuzz before then since it did not exist, so it does not make sense to evaluate Syzbot on this time.
Furthermore, SyzRetrospector is unable to faithfully recreate Syzbot's environment before it existed, so any venture there would be invalid.
In this sense, the beginning of Syzbot acts as a revealing factor for bugs that pre-date it and are not otherwise hidden.
Since we do not know the exact date Syzbot was turned on, we will use the date of the first bug found by it: July 22nd, 2017.
$D_2$ is the delay while the bug is revealed, but the fuzzer has not found it yet (\textit{Finding Date} - \textit{Revealing Date}).
The bugs in our pilot study have an overall delay ($D_1 + D_2$) of $405.11$ days, but the bugs were only revealed for an average of $73.94$ days ($D_2$).
This delay is the actual time it takes for the fuzzer to find the bugs.
The remaining $331.17$ days ($D_1$) is the delay for the fuzzing environment to improve.

\subsection{Revealing Factors}
\label{subsec:revealing_factors}

Here we provide a list containing each of the possible revealing factors for a bug, as well as real-world examples of each.

\subsubsection{Syscall Description Update}
Syscall descriptions are manually written and built into Syzkaller to allow it to generate test programs which call syscalls.
Bugs are often hidden behind syscall descriptions that have not yet been implemented into Syzkaller.
Consider a memory leak in \texttt{kobject\_set\_name\_vargs}~\cite{pilotbug4}.
In this case, the bug is not in the \texttt{kobject} library, but a misuse of it in the \texttt{nilfs} filesystem.
After deleting a \texttt{kobject} using \texttt{kobject\_del}, the calling function must call \texttt{kobject\_put} in order to free the object.
Not doing so constitutes a memory leak.
This bug was introduced on August 8th, 2014, nearly $3$ years before Syzbot began fuzzing.
However, once Syzbot began, it still lacked the proper syscall descriptions to fuzz the \texttt{nilfs} module.
Support for this filesystem was added along with support for $19$ other filesystems on September 20th, 2020.
With the new descriptions built in, Syzbot was able to find the bug $57$ days later on November 16th, 2020.

For bugs revealed this way, it is important to note that \textit{hidden} refers to Syzbot's ability to find the bug.
Syzbot could not find the bug because it lacked the interface to fuzz the vulnerable module.
A bug like this may very well be exploitable before it is revealed, as well as after; a 6 year window in this case.

\subsubsection{Kernel Commit}
\label{subsubsec:kernel_commit}
There are some cases where the root cause of a bug and the kernel commit that reveals it are not the same.
By our definitions: the guilty commit is the commit that introduces the root cause of a bug, while the revealing commit reveals an already present flaw to the fuzzer.
This can be seen clearly in the following bug: \texttt{WARNING in rtl28xxu\_ctrl\_msg/usb\_submit\_urb}~\cite{pilotbug1}.
The bug arises when the message pipe directions do not match between host and USB device, which can happen for some zero-length control requests.
This behavior had existed since February 3rd, 2015, but was not findable by Syzbot since it would not generate a crash; the control request would simply fail.
On May 22nd, 2021, a warning was added to check the pipe direction before handling the request, and thus the bug was revealed.
Syzbot found the bug $2$ days after the revealing commit was pushed.
Added warnings and checks are the easiest reveal to explain, but there are plenty of others, such as an additional function call that completes the code path to the crash site~\cite{casestudybug5}.
In general, the kernel is already in a buggy state, and the revealing kernel commit turns this state into a crash.

\subsubsection{Sanitizer Commit}
A sanitizer commit is a commit that specifically introduces or improves a sanitizer.
Since sanitizers are built into the kernel, it is a type of kernel commit.
Despite this, sanitizers are developed independently of the kernel and are not intended to be built into consumer releases such as Ubuntu.
So, we consider them to be a separate revealing factor from other kernel commits.
We note that there were no bugs revealed by sanitizer commits in our pilot study.
However, since sanitizers play such a huge role in finding bugs, it stands to reason that their development could reveal bugs.
We will explain this revealing factor's rarity in \S\ref{subsubsec:easy_reveals}.

\subsubsection{Blocking Bug}
\label{subsubsec:blocking_bugs}
A blocking bug is a bug in the Linux kernel that somehow prevents Syzbot from finding another bug.
It is often a crash earlier in the call stack of the hidden bug that triggers most of the time, thus preventing the fuzzer from continuing.
A good example is seen in the bug \texttt{WARNING in exception\_type}~\cite{pilotbug9}.
This bug occurs in KVM when the guest's maximum physical address (\texttt{MAXPHYADDR}) is set to 1.
The warning is already difficult to hit due to the absurd nature of the state.
KVM prevents \texttt{MAXPHYADDR} from going below 32 in the hypervisor, though the buggy value still exists in the guest's memory.
So the warning is only triggered if the guest's memory is checked directly.
This does not prevent the bug, it only makes it harder to find.
On top of that, there is a much more common bug that Syzbot finds in KVM: \texttt{WARNING in x86\_emulate\_instruction}.
So, rather than finding the warning in \texttt{exception\_type}, which is already hard to find, it finds the blocking bug.
The blocking bug was patched on May 28th, 2021, and the warning in \texttt{exception\_type} was found on August 29th, 2021.

We observe that blocking bugs tend to block probabilistically.
That is, while the blocking bug exists, it is not impossible to find the hidden bug, just infeasible.
So, the revealing commit of a bug hidden by a blocking bug may not be the patch for that blocking bug.
However, we will demonstrate that this does not affect the accuracy of our results in \S\ref{subsec:case_study}.

\subsubsection{Syzkaller Commit}
Bugs are sometimes revealed by changes to Syzkaller itself.
This often occurs when Syzbot gains additional hardware or software support, or changes its fuzzing method for a specific module.
In the case of \texttt{KASAN: slab-out-of-bounds Read in packet\_recvmsg}~\cite{pilotbug12}, Syzbot was already capable of fuzzing the buggy module (WireGuard), but needed assistance in setting up initial devices.
Setting up complex scenarios can be difficult for fuzzers as their input is entirely random and may never create something as complex as a valid network.
A Syzkaller commit on February 13th, 2020 changed how Syzkaller fuzzes WireGuard by initializing its virtual network with 3 devices.
With this support, Syzbot was able to find the bug on March 12th, 2022.

\subsubsection{Never Hidden}
Some bugs are never hidden from Syzbot and can be reproduced at their guilty commit.
The bug \texttt{WARNING in futex\_requeue}~\cite{pilotbug10} occurs as the result of a race between the function \texttt{futex\_requeue} and the top waiter in the \texttt{futex} queue.
If the waiter wakes up and exits before \texttt{futex\_requeue} finishes, the warning is triggered.
There is a 13 day gap between when the bug was introduced and when it was found.
Further, our pilot study shows that Syzbot was capable of finding the race the whole time.

\begin{figure*}[t]
    \centering
    \includegraphics[width=0.85\textwidth]{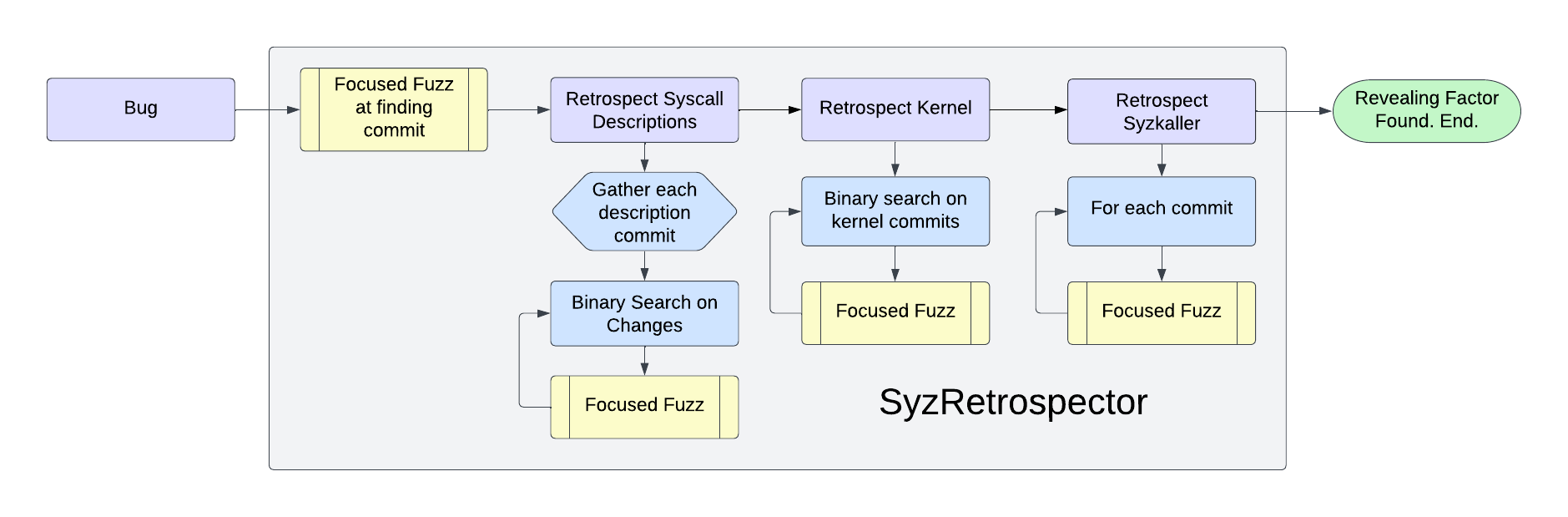}
    \vspace{-0.2in}
    \caption{SyzRetrospector Workflow.}
    \vspace{-0.25in}
    \label{fig:workflow}
\end{figure*}

\section{Overview}
\label{sec:overview}

Building on our pilot study, we construct SyzRetrospector, an analysis tool capable of going back in time to faithfully recreate Syzbot's fuzzing environment at any point in time.
In this section, we provide a high-level overview of SyzRetrospector's workflow (Figure~\ref{fig:workflow}) and how it goes back in time to narrow down commits to a single revealing commit.
In this section and next, we will discuss various elements shown in this figure.

In order to determine whether a bug is findable in a commit, SyzRetrospector faithfully recreates Syzbot's fuzzing environment on the day of that commit and then fuzzes for the bug.
Syzbot's fuzzing environment is comprised of \textit{a kernel commit, a Syzkaller commit, and a set of syscall descriptions}.
The syscall descriptions are taken from a Syzkaller commit which we refer to as a syscall description commit.
Each of our $5$ revealing factors is one of these types of commits.
We analyze each of these groups in turn, but in the context of the other groups.
So, if SyzRetrospector is testing a syscall description commit, it chooses Syzkaller and Linux commits from the same day to complete the environment.
This means that as it switches from one commit type to the next, the shrinking date range of where the revealing factor could be carries over.

Given any bug, SyzRetrospector starts by fuzzing at that bug's finding commit, where we know the bug exists.
Fuzzing here tells SyzRetrospector whether the bug is findable in a timely manner, and how long the bug takes to reproduce on average.
Bugs that are not found, often hard-to-find races, are skipped, while the rest move on to retrospection.
We will justify this decision in our threats to validity (\S\ref{subsec:unfound_bugs}).

\begin{figure*}[t]
  \centering
  \vspace{-0.1in}
  \includegraphics[width=0.85\textwidth]{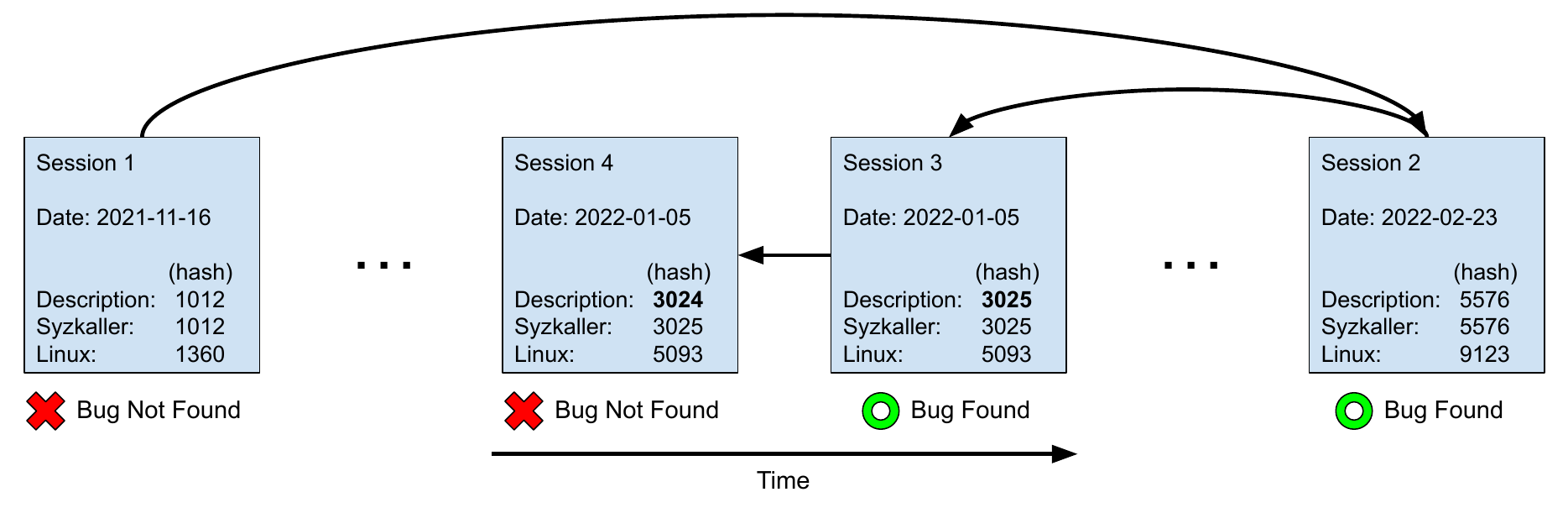}
  \vspace{-0.1in}
  \caption{SyzRetrospector determines which commit revealed the bug in question as the bug reproduces after, but not before, the revealing commit. Here that is the syscall description commit with hash $3025$. The example hashes are monotonically increasing.}
  \vspace{-0.2in}
  \label{fig:commit_confirm}
\end{figure*}

First, SyzRetrospector gathers and retrospects all of the relevant syscall description commits.
It does this by parsing the bug's reproducer and the Syzkaller git to see which commits changed or introduced the syscall descriptions in it.
SyzRetrospector then uses binary search to find the earliest syscall description commit that is able to reproduce the bug, and confirms whether this commit is the revealing factor by using the same version of Linux and Syzkaller, but an older description set.
This process is depicted in Figure~\ref{fig:commit_confirm}.
If the bug is found after the description commit, but not before, the description commit must be the revealing factor.
This provides a simple algorithm to isolate and then confirm whether a syscall description commit revealed the bug in question.
Otherwise, the revealing factor is either a Linux or Syzkaller commit between the dates of this description commit and the one before it.

Second, SyzRetrospector searches the kernel commits for a revealing factor.
Again, it uses binary search to narrow down the commits.
And once again, if the bug is found after a kernel commit, but not before, then that commit is the revealing factor.
Otherwise, the revealing commit must be a Syzkaller commit.
Since kernel commits are pushed upstream daily, the date range has shrunk to a single day.

Third, there are relatively few Syzkaller commits each day, so SyzRetrospector searches them linearly going back in time.
Once the bug reproduces after, but not before a commit, that commit must be the revealing factor.

\section{Design}
\label{sec:design}

SyzRetrospector is purpose-built to identify the revealing factors of bugs and thus their $D_1$ and $D_2$.
To accomplish this, SyzRetrospector looks back in time, building old versions of both Syzkaller and the Linux kernel.
It is even able to determine the correct C compiler based on what Syzbot has used in the past.
In this section, we will explain in more detail the workflow laid out in \S\ref{sec:overview}, including how we focus Syzkaller to a specific set of syscalls and how it arrives at a correct result.

\subsection{Preliminary Fuzzing Session}
\label{subsec:preliminary_session}

Before retrospection begins, SyzRetrospector gathers everything it needs related to the bug from Syzbot, including the bug's reproducer, its finding and guilty commits, and the exact kernel configuration used by Syzbot while fuzzing.
It also makes note of any bugs which share the same patch as these are duplicates, or different manifestations of the same bug.
They are treated as the same bug during retrospection.

As demonstrated in Figure~\ref{fig:workflow}, SyzRetrospector begins by attempting to generate the bug at the finding commit while using the last Syzkaller commit from the same day.
In general, if an arbitrary commit is needed for a specific day, the last commit of that day is chosen as representative of that day.

For this and all other fuzzing sessions, SyzRetrospector begins by cutting down the number of syscall descriptions to only those needed to find the bug.
If we were to let Syzkaller run without any form of guidance, it would simply try to maximize coverage and may never find the bug we want it to.
At the same time, we do not want to change how Syzkaller fuzzes as that impacts our results.
Instead, we will restrict what areas of code Syzkaller can fuzz by focusing the syscall descriptions built into it.
We call this contribution \textit{focused fuzzing}, and it is one of the core reasons SyzRetrospector is able to scale to a large number of bugs with relatively limited compute power.
This focused fuzzing scheme greatly reduces the time taken to find a single bug from many days to only a few hours.
To enable focused fuzzing, SyzRetrospector parses the large, unfocused description set that Syzkaller would pull from randomly, and narrows it down to only the descriptions and dependencies required to reproduce the bug.
The exact process and challenges involved in this are described in \S\ref{subsec:desc_parsing}.
After the syscall descriptions have been focused, Syzkaller has a fully functional subset of the original descriptions, usually around $20$ syscall descriptions.
Importantly, none of the descriptions or underlying structures are changed.
We have only limited the ones Syzkaller is allowed to use.

We further decrease the time to find by inserting the reproducers into the corpus as seeds.
By encouraging these test cases, Syzkaller finds most bugs in $2$ minutes.
Importantly, we leave Syzkaller's mutation and scheduling algorithms alone.
This way, we encourage Syzkaller to use the reproducers and similar inputs, but still allow the fuzzer to generate and mutate test cases as it normally would.
Since mutation is left untouched, Syzkaller is still capable of finding new paths to the same bug if the reproducer fails to trigger it.
We reason that using the reproducer does not interfere with SyzRetrospector's faithful recreation of Syzbot's fuzzing environment since it does not change how the fuzzer operates.

Next, the kernel is built.
SyzRetrospector downloads and uses the same configuration that Syzbot used when it found the bug.
Then, multiplexing the set of compilers that Syzbot has used, SyzRetrospector chooses a suitable compiler based on the date of the kernel commit (\S\ref{subsec:compiler_selection}).
This ensures the kernel builds without error.

\subsubsection*{Maximumm Fuzzing Time}
\label{subsubsec:fuzzing_time}
As Syzretrospector searches for the bug in question, it needs to know when to stop searching.
At the finding commit, SyzRetrospector fuzzes $3$ times for $30$ minutes in order to understand how long it takes for the bug to reproduce.
It then recalculates the maximum fuzzing time by using the mean $+$ standard deviation of the 3 trials, which can end up being greater than 30 minutes.
For most bugs, the new time is around $10$ minutes (a minimum set by us), and the highest is around $33$ minutes.
SyzRetrospector uses this time going forward except in two cases.
First, if any bug takes $80\%$ of its maximum time to reproduce, SyzRetrospector automatically resets the maximum time to $30$ minutes from what was previously calculated, and fuzzes $5$ times rather than $3$.
This is in an attempt to catch bugs that are harder to find than we originally thought.
Second, we noticed that memory leaks are notorious for this kind of behavior, so SyzRetrospector will always use 5 attempts at $30$ minutes for these types of bugs.
Calculating the maximum time in this manner further decreases SyzRetrospector's overall run time by letting it better decide when a bug is not findable.
From our experience, the time needed to find a bug does not change
as we move back in time.
Thus the maximum fuzzing time is not recalculated past the preliminary session.
Bugs that are found at the finding commit proceed to the rest of the retrospection algorithm as it is covered in \S\ref{sec:overview}.

\subsection{Result Collection}
\label{subsec:result_collection}

Once retrospection is completed, SyzRetrospector outputs a final report with a log of all the fuzzing sessions, the exact revealing commit, and important dates in the bug's lifetime.
However, some manual work is still required to classify the revealing factors.
SyzRetrospector differentiates between syscall description and Syzkaller commits, so they can be taken at face value.
Kernel commits, however, could represent a sanitizer commit, a blocking bug, a regular kernel commit, or the guilty commit.
Bugs that are never hidden (\ie are findable at their guilty commit) are easy to rule out and this is even done automatically.
Similarly, sanitizer commits can be recognized by their commit name, but this process is still manual.
Otherwise, SyzRetrospector automatically reports any bugs it thinks may be blocking the bug in question.
These are bugs that appear many times throughout retrospection, especially when the original bug is not found.
The duty still falls to the researcher to look through the log and determine if the bug is a blocking bug or not.
Finally, if all other options are ruled out, then the kernel commit is what revealed the bug.

\section{Implementation}
\label{sec:implementation}
We faced and solved many challenges to successfully realize SyzRetrospector, which is comprised of $7806$ lines of C++ code and BASH script.
Many of these challenges stem from looking back in time to fuzz many commits, and matching the scale of Syzbot.

\subsection{Git Traversal}
\label{subsec:git_traversal}

At first glance, a git repository
is a simple tree where commits are added and merged into one child, and a path to a parent is as simple as walking the tree out to it.
Even better, the repository might already be a simple line, one commit after the next.
For Linux, this could not be further from the truth.
Merges and branches occur haphazardly and plentifully even within a single repository.
Because the repositories can be so complicated, SyzRetrospector cannot just clone the kernel git and search between the finding and guilty commits.
On top of that, we want to use binary search to make SyzRetrospector more efficient.
In order to properly binary search on the kernel commits, SyzRetrospector must draw a single line of commits from finding to guilty.
As it parses back in time, we want SyzRetrospector to remain on the upstream branch to avoid drastic code changes from switching branches.
So, when a merge commit is reached, SyzRetrospector chooses the first parent to follow.
The first parent of a merge commit is always the branch that was merged into, so SyzRetrospector remains on the same branch: upstream.

The Syzkaller git is much simpler by comparison, with only a single chain of commits which can be searched as needed.

\subsection{Duplicate Bugs}
Duplicate bugs can be difficult to identify automatically, even though SyzRetrospector has some simple deduplication built into it.
When gathering a bug from Syzbot, it will check for any other bugs with the same patch as they share the same root cause, and are thus the same bug.
However, some duplicates are not yet found by Syzbot.
During analysis, if SyzRetrospector found bugs that appear to be duplicates, we manually study them and mark them as such.
We use heuristics such as similar stack trace, same crashing function, same crash type, same sanitizer, etc. to consistently mark duplicate bugs.

\definecolor{dkgreen}{rgb}{0,0.6,0}
\definecolor{mauve}{rgb}{0.58,0,0.82}

\lstset{frame=tb,
  language=C,
  keywords={},
  showstringspaces=false,
  columns=flexible,
  basicstyle={\scriptsize\ttfamily},
  numbers=none,
  keywordstyle=\color{blue}\bf\ttfamily,
  commentstyle=\color{dkgreen},
  stringstyle=\color{mauve},
  breaklines=true,
  breakatwhitespace=true,
  tabsize=3,
  moredelim=**[is][\color{blue}\bf\ttfamily]{@}{@},
  moredelim=**[is][\color{dkgreen}\it\ttfamily]{^}{^}
}

\begin{figure}
    \centering
    \scriptsize
    \vspace{0.1in}
    \begin{minipage}[b]{\columnwidth}
    \begin{lstlisting}
     1   resource @my_resource@[intptr]
     2
     3   @producer@(num int32) my_resource
     4   @consumer@(p ptr[inout, parent])
     5
     6   @parent@ {
     7      ...
     8      child    child    (in)
     9   }
     10
     11  @child@ {
     12     ...
     13     rec      my_resource
     14  }
    \end{lstlisting}
    \end{minipage}
    \caption{An example syscall description}
    \label{fig:template_code}
    \vspace{-0.2in}
\end{figure}

\subsection{Syscall Description Parsing}
\label{subsec:desc_parsing}

In order to achieve focused fuzzing as described in \S\ref{subsec:preliminary_session}, SyzRetrospector needs to be able to parse the syscall descriptions of any Syzkaller commit.
While more recent versions of Syzkaller allow a user to enable only a subset of syscall descriptions, older versions do not have this feature.
SyzRetrospector must carve out a new description set for each bug, and for each version of the descriptions.

Syzlang, the custom language of the syscall descriptions, is relatively straightforward to parse thanks to the rigorous and consistent writing of the Syzkaller team.
Each syscall depends on a number of structures, unions, types, resources, definitions, and flags that describe how data should be passed with the syscall.
In Figure~\ref{fig:template_code}, the syscall \texttt{consumer} depends on the structure \texttt{parent}, which in turn depends on the structure \texttt{child}.
These are easy enough to find and add to the new description, keeping in mind that those items may depend on yet more items.
Resources, however, pose a greater challenge.

Each resource must be produced, or provided as output from a syscall in the description, and consumed, or used as an input by another syscall, otherwise Syzkaller will not build.
The nuance arises because resources are not always easily identified as input or output.
A pointer in a syscall may be labeled as \texttt{in} or \texttt{out} or \texttt{inout} (both in and out)~\cite{syzbot_syzlang}.
Items pointed to by that pointer will be used as input, output, or both, which must be accounted for.
In the example, \texttt{parent} is passed to \texttt{consumer} as an \texttt{inout} pointer, \texttt{parent} houses the structure \texttt{child} as an \texttt{in} (input) structure, and finally \texttt{child} has a member \texttt{my\_resource}.
So \texttt{consumer} uses \texttt{my\_resource} as an input.
A resource could also be passed as an argument to a type template, making it harder to track.

Following the rules above, SyzRetrospector begins by parsing all of the syscall descriptions from Syzkaller and classifying each object.
It then consults the reproducer to determine which syscall descriptions are required and adds those to a new description.
SyzRetrospector then enters a loop of adding the dependencies of the objects already in the new set.
If one of those dependencies is a resource, it adds up to two new syscalls: one that produces it, and one that consumes it.
While choosing these syscalls, SyzRetrospector tries to select those that do not depend on other resources to reduce the total number of syscalls in the end.
Lastly, Syzkaller always needs a few specific syscall descriptions in order to function.
Through our testing, we found that Syzkaller breaks at runtime without \texttt{mmap} and \texttt{syz\_execute\_func}.

Thankfully, the language of Syzkaller's syscall descriptions, Syzlang, largely remains unchanged through the past $6$ years, and most changes are simply additions of features.
The only functional change we have had to account for is the meaning of \texttt{inout} as an object attribute.
This keyword currently means that children objects, specifically resources, will be labeled as either \texttt{in} or \texttt{out} as needed.
This nuance allows the developers to have members of the same structure function as either producers or consumers in a single syscall.
However, in older versions of Syzkaller, \texttt{inout} meant that every child object should be treated as both in and out.
In this case, the description parser needs to gather all children objects, regardless of attribute.

\subsection{Patches}
\label{subsec:patches}

SyzRetrospector goes back in time to look at Syzkaller and Linux commits alike.
However, large swathes of commits in both Linux and Syzkaller either do not build or do not boot.
In order to obtain accurate retrospection, we have written a series of patches for both the kernel and Syzkaller.
Through manual effort, we have denoted exact commit ranges to which each commit needs to be applied.

To avoid affecting the bugs we are trying to reproduce, we patched the kernel as little as possible.
For example, many boot errors could be solved by changing some of Syzkaller's boot parameters.
Other build errors could be fixed by selecting the right compiler (\S\ref{subsec:compiler_selection}).
Still, some errors could only be fixed by manually patching the kernel.
One particular bug had the kernel building itself with the incorrect page size, and then breaking on boot.
Our patch is based off of the one used in mainline Linux and simply forces a 2 MB size page~\cite{kernelpatch1}.
SyzRetrospector has $7$ similar patches that it checks before building any instance of the kernel.

We also develop and maintain $9$ patches for Syzkaller that either build Syzkaller correctly, or ensure that Syzkaller boots Linux properly.
For example, QEMU needed an extra argument appended to it, \texttt{-cpu host,migratable=off}, before it was added on October 28, 2018.
But then between May 1st, 2020 and January 1st, 2021, that same argument needed to be shortened to just \texttt{-cpu host}.
In total, there are $16$ patches SyzRetrospector considers before building Syzkaller.

While we tried to keep these patches simple and non-invasive, they often dealt with core components such as memory.
Indeed, the very nature of patching implies that we are removing bugs from the kernel.
We reason that if left unpatched, thousands of commits would be unusable.
It is better for us to patch the errors in the likelihood that other bugs are not affected.
We also believe that these patches do not introduce more bugs as we did not find any during our study.

\subsection{Unstable Commits}
\label{subsec:unstable}

The patches in the previous section are a key part of allowing SyzRetrospector to function over wide ranges of Linux and Syzkaller commits.
However, it is infeasible to patch every boot error or incompatibility.
Instead, we implement SyzRetrospector with the ability to work around errors.

SyzRetrospector is able to detect issues in two ways: it can detect boot failures by reading the Syzkaller logs, and it can detect incompatibility crashes by parsing the crashes.
For instance, a crash with the name ``lost connection to test machine'' signifies a crash that stopped the kernel so abruptly that Syzkaller could not generate a crash report.
Similarly, any crash beginning with ``SYZFAIL'' or ``SYZFATAL'' represents a failure of some kind in the fuzzer.
If SyzRetrospector identifies an issue, it marks the set of kernel commit, Syzkaller commit, and syscall descriptions as unstable.
SyzRetrospector continues retrospection, but attempts to work around the unstable commits by fuzzing nearby commits.
In the case that the revealing factor is inside a range of unstable commits, SyzRetrospector reports this in the final report.
In most cases, SyzRetrospector successfully works around the unstable commits and identifies the revealing factor outside of them.
We observe that if SyzRetrospector was unable to work around unstable commits, it would incorrectly assume that a bug was unfindable, negatively impacting our results.
We will demonstrate that this impact has been mitigated in \S\ref{subsec:case_study}.

\subsection{Compiler Selection}
\label{subsec:compiler_selection}

In addition to choosing Linux commits to fuzz, SyzRetrospector is able to select the correct version of the C compiler GCC.
Changing the compiler can make a huge difference in whether a bug is findable, or whether the kernel even builds.
In some cases, compile warnings in older GCC versions are treated as errors in newer versions.
Some older kernel versions do not comply with the rules that were made after them, so we need to use older GCC versions just to get the kernel to compile.
Also, the GCC version used by Syzbot naturally changed over time.
We have looked back on the history of GCC compilers used by Syzbot and compiled a collection of them to be multiplexed by SyzRetrospector.
We choose the compiler based on the date of the kernel commit we want to build, and compare that to the compiler Syzbot used during the same time frame.

\subsection{Architecture}

In addition to the x86\_64 architecture, Syzbot also fuzzes i386, using cross-compiled test cases.
In these cases, kernel compilation does not change, but parts of Syzkaller must be cross-compiled.
To allow for this cross-compilation, we make use of syz-env~\cite{syz-env}, a docker container used to cross-compile Syzkaller.
We note Syzbot has a spread of 6 managers that fuzz RISC-V and Arm 32 and 64 bit versions.
However, these managers only fuzz auxiliary repositories, so their crash instances are left out of our study.
If their bugs are found in upstream, the upstream crash is used for retrospection.

\subsection{Optimizations}

In order to further improve the runtime of SyzRetrospector, we made several small optimizations to it.
First, for each fuzzing session, once the bug is found, there is no need to continue fuzzing.
SyzRetrospector stops fuzzing before the maximum time is reached, and moves on to the next session.
Second, SyzRetrospector keeps track of the sessions that we have fuzzed in the past.
If a bug was previously tested on a specific combination of kernel commit, syscall description set, and syzkaller commit, we can simply use the results from the previous session.
This case is not common, but occurs when disambiguating the result of binary search.

\section{Data Set}
\label{sec:data_set}

Each of the $559$ bugs in our data set meets strict requirements to ensure valid results.
We gathered the bugs for retrospection on February 7th, 2023, so many newly fixed bugs are not included.
First, Every bug needs a fixing commit as we get its guilty commit from the fixing commit's \texttt{Fixes} tag.
Some bugs have multiple fixing commits, such as when the fix was incomplete.
If they do, we look into each to find guilty commits.
If a bug has multiple guilty commits, we take the oldest to be the guilty date.
It then needs a reproducer, not only so we know what syscall descriptions to use, but so the bug will reproduce faster.
Since we have found erroneous cases where a bug was found before its guilty commit or found after it was marked as fixed, we also perform a sanity check that all of the dates for a bug are in order.
We also found that the auxiliary repositories such as \texttt{linux-next} often undergo major changes, such as deleting branches or resetting the repository, leaving many commits dangling.
These commits are often the guilty and/or finding commits that SyzRetrospector needs to study a bug.
Due to the transient nature of these repositories, we only study bugs that appear upstream.
As we will discuss in \S\ref{subsec:repo_choice}, we believe this choice is acceptable.
We gathered a total of $1159$ bugs for SyzRetrospector to study.
However, many of these bugs failed to reproduce even with their reproducer, and others encountered build errors that remain unsolved.
As we justify in \S\ref{subsec:unfound_bugs}, these bugs were left out of the study.

\section{Findings}
\label{sec:findings}

Here we present our findings after performing a large scale retrospection of $559$ bugs found by Syzbot.
This process took over $3.5$ months and over $10,170$ CPU hours.

\subsection{50 Bug Case Study}
\label{subsec:case_study}

The non-deterministic nature of fuzzing and many of the bugs we retrospected present an interesting challenge for us, as bugs may not always reproduce as we expect them to.
This results in false negatives for individual fuzzing sessions, and possibly incorrect results overall.
In order to ensure the accuracy of SyzRetrospector, we undertook a case study of 50 bugs chosen randomly from the data set.
In this case study, we carefully inspected the revealing factors and dates given by SyzRetrospector in order to verify their correctness.
In addition, we manually looked for other possible revealing factors.

Out of the 50 bugs studied, we found that $44$ of them had correctly identified revealing factors.
So, SyzRetrospector reaches the correct result $88\%$ of the time.
The bugs that did fail were hard-to-find bugs that managed to reproduce at their finding commits, so they were included in the study.
As the bugs did not reproduce consistently, they resulted in false negatives.
On average, the incorrect results were off by $19$ days, with one outlier at $119$ days, compared to the average $D_1$ and $D_2$ which are $331.17$ and $73.94$ days respectively.
We argue that the error is small enough that SyzRetrospector's overall results are reliable.

\begin{figure}
    \begin{minipage}[l]{\linewidth}
        \centering
        \includegraphics[width=\columnwidth]{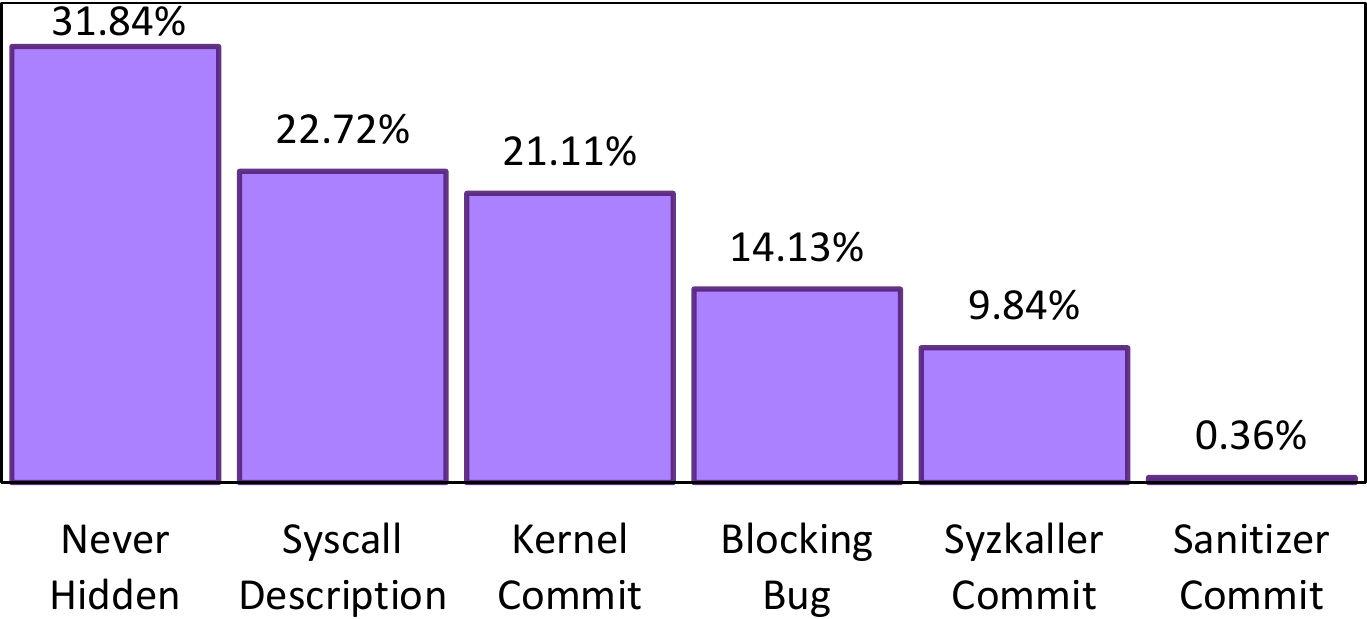}
        \caption{Percent of bugs revealed by each revealing factor.}
    \label{fig:revealing_factors}
    \end{minipage}
    \hspace{0.1in}
    \begin{minipage}[r]{\linewidth}
        \centering
        \includegraphics[width=\columnwidth]{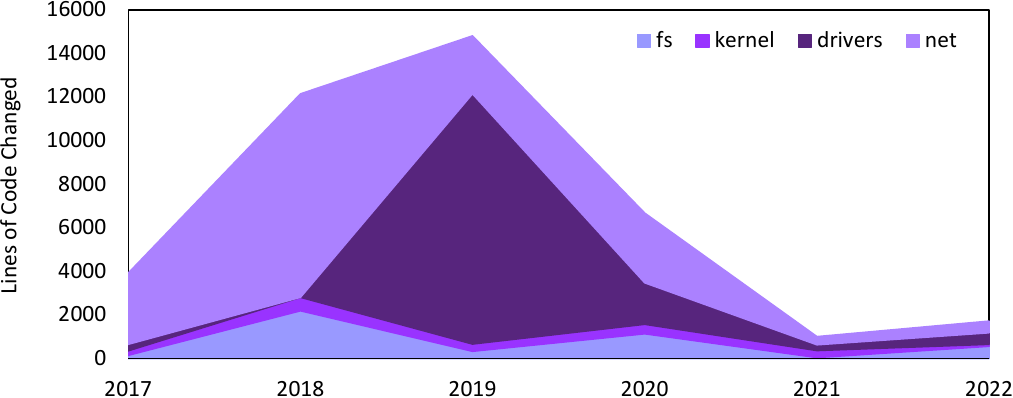}
        \caption{Lines of code changed in syscall descriptions each year.}
        \label{fig:template_churn}
    \end{minipage}
    \vspace{-0.2in}
\end{figure}

\subsection{Revealing Factors}
\label{subsec:rf_spread}

Figure~\ref{fig:revealing_factors} shows the revealing factors and the percent of bugs they are responsible for revealing.
We see $68.16\%$ of bugs were hidden for some portion of their lives, the most prominent categories being syscall description commits, kernel commits, and blocking bugs.
We note that each of these revealing factors requires developer work to induce the reveal, but that each category is unique in the work required.
Moreover, some revealing factors are much harder to induce than others.

\subsubsection{Hard-to-Induce Reveals}
Among the 5 reveal groups, kernel commits and blocking bugs are the hardest to induce.
Blocking bugs, for instance, could be any of the open bugs in Linux.
A developer has no way of knowing which of the open bugs are blocking bugs before they are patched.
This creates a circular dependency where in order to know which bugs need to be patched, one would have to patch those bugs.
The only way to improve $D_1$ for these bugs is to speed up the patching process, which is challenging due to the large number of players involved.

Kernel commit reveals are difficult to predict since it is unknown what commit needs to be written to reveal a bug.
Consider a slab-out-of-bounds bug in the Thrustmaster joystick driver~\cite{unexpectbug111}.
The driver holds a pointer to a USB endpoint that has already been freed, but the driver fails to check the endpoint before attempting to use it.
The bug was hidden for a long time since even though the pointer enters a buggy state, it usually does not point out of bounds.
We now shift to the revealing factor: a commit in the B-Tree Filesystem (BTRFS) module\cite{unexpectreveal111}.
Periodically, BTRFS compresses inodes to save memory, but prior to this commit, did not compress single-page inodes.
The Thrustmaster pointer happens to point into one of these single-page inodes.
Once the inode is compressed, the Thrustmaster pointer, which uses fixed offsets, now points outside the valid memory range, and can trigger Kasan.
Many kernel commit reveals are similar in that it is hard to predict what change will reveal a bug.
However, we will provide suggestions on how to improve the chances of inducing kernel commit reveals in \S\ref{subsec:improve_d1}.

\begin{tcolorbox}
    \textbf{Finding \nextfnum.}~~
    We find that $35.24\%$ of bugs are hidden by kernel commits and blocking bugs, and that it is hard to improve the $D_1$ of these bugs.
\end{tcolorbox}

\subsubsection{Easier-to-Induce Reveals}
\label{subsubsec:easy_reveals}
Contrary to the reveals listed above, syscall description commits, Syzkaller commits, and sanitizer commits have more predictable outcomes.
Table~\ref{tab:reveal_by_year} shows the number of reveals of each type spread out by year.
This reveals a surprising trend.
Despite most of the syscall description reveals occurring in 2020, Figure~\ref{fig:template_churn} shows more changes to the description set in 2019.
Since 2020 had more revealed bugs from fewer lines changed, it must have had smaller, more impactful commits.
By our observation, one such commit adds support for 20 filesystems~\cite{syzkallercommit1} that revealed many bugs.
Based on this, it would appear that not all syscall descriptions are created equal.
Some description commits, despite being small, are able to reveal more bugs than larger ones.

\begin{table}
    \centering
    \scriptsize
    \caption{The number of reveals of each type by year.}
    \begin{tabular}{rcccccc} \toprule
        Year & \makecell{Description \\ Commit} & \makecell{Syzkaller \\ Commit} & \makecell{Kernel \\ Commit} & \makecell{Blocking \\ Bug} & \makecell{Sanitizer \\ Commit} & \makecell{Never \\ Hidden}\\ \midrule
        2017 & 5 & 6 & 1 & 3 & 0 & 2\\
        2018 & 23 & 15 & 28 & 11 & 0 & 28\\
        2019 & 21 & 12 & 22 & 21 & 2 & 38\\
        2020 & 59 & 12 & 25 & 20 & 0 & 49\\
        2021 & 15 & 9 & 21 & 11 & 0 & 30\\
        2022 & 4 & 1 & 21 & 13 & 0 & 31\\
        \bottomrule
    \end{tabular}
    \vspace{-0.1in}
    \label{tab:reveal_by_year}
\end{table}

Table~\ref{tab:reveal_by_year} also shows that in recent years, fewer bugs have been revealed by Syzkaller commits and syscall description commits.
In fact in 2022, only 5 bugs were revealed by improvements to Syzkaller.
This correlates strongly with the decline of syscall description development shown in Figure~\ref{fig:template_churn} in the years 2021 and 2022.
At the same time, a higher percentage of bugs are findable from their guilty commits and the number of bugs found by Syzbot has not dropped~\cite{syzbot}.
This means that the commits to Syzkaller in previous years continue to pay off as they allow Syzbot to find many bugs.
This is a good sign for Syzbot going forward.
On the other hand, had the development of syscall descriptions continued, Syzbot would have likely revealed many bugs that otherwise remained hidden in those years.
In fact, \S\ref{subsec:improve_d1} will show specific areas that require more development.

\begin{tcolorbox}
    \textbf{Finding \nextfnum.}~~
    We find that Syzbot continues to find bugs based on past improvements to its description set, but that more development is likely to reveal more bugs.
\end{tcolorbox}

Sanitizers are an interesting revealing factor since it appears they do not reveal many bugs.
Despite $42.75\%$ of bugs in our data being found with the assistance of sanitizers, only $2$ were actually revealed by commits to sanitizers.
A deeper look into the sanitizers' development reveals what is actually going on.
Take KASAN for example, one of the first sanitizers in Linux.
It has only seen around $415$ commits since it was introduced in 2015.
Furthermore, almost all of the commits since 2019 are minor fixes and selftests.
It makes sense that none of these commits would reveal many, if any, bugs.

KMSAN, however, is unique in that it was introduced to an already mature Syzbot.
We are able to use this introduction to show that new sanitizers do reveal bugs when they are added to the kernel.
We manually studied the behavior of $8$ bugs found by KMSAN around the time it was introduced.
We traced $4/8$ of the bugs back to the commit that added KMSAN.
$3$ of the bugs were revealed some time after KMSAN was added, and $1$ was blocked by a 9th KMSAN bug.
This 9th bug is not included since Syzbot has marked it as invalid~\cite{kmsanbug9}.
So, reveals induced by sanitizers are predictable.
Adding a new sanitizer will reveal several bugs as well as allow the fuzzer to find an entirely new class of bugs, but the little development that occurs after does not reveal many bugs.

\begin{tcolorbox}
    \textbf{Finding \nextfnum.}~~
    We find that new sanitizers bring a jump in fuzzing capability, revealing several bugs. However, the little development after they are introduced does not reveal many bugs.
\end{tcolorbox}

\begin{figure*}[t]
    \centering
    \begin{subfigure}[b]{0.31\textwidth}
        \includegraphics[width=\linewidth]{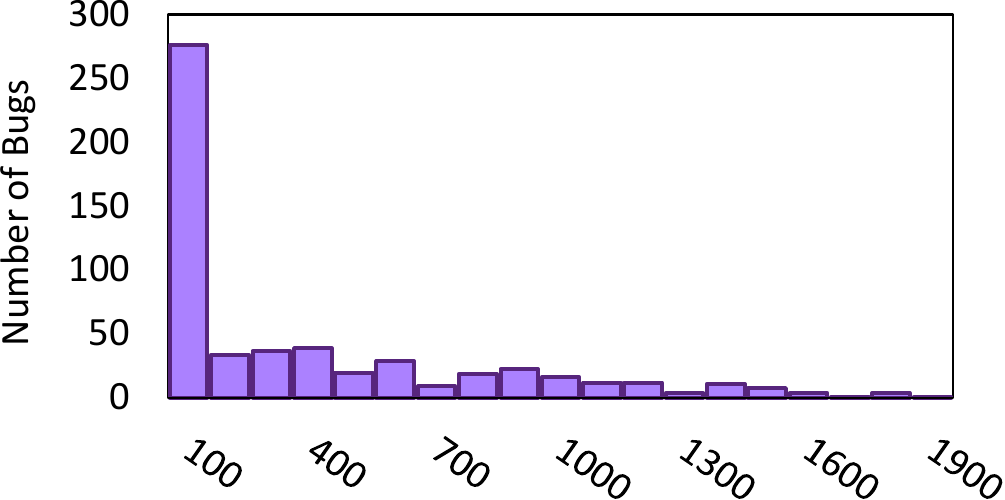}
        \caption{The distribution of $D_1$.}
        \label{fig:d1_hist}
    \end{subfigure}
    \begin{subfigure}[b]{0.27\textwidth}
        \vspace{-0.15in}
        \includegraphics[width=\linewidth]{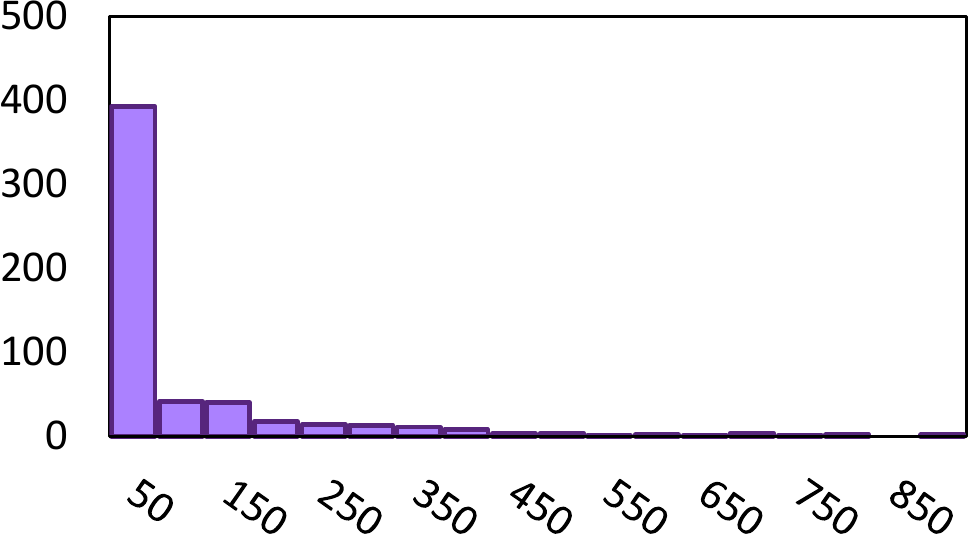}
        \caption{The distribution of $D_2$.}
        \label{fig:d2_hist}
    \end{subfigure}
    \begin{subfigure}[b]{0.28\textwidth}
        \includegraphics[width=\linewidth]{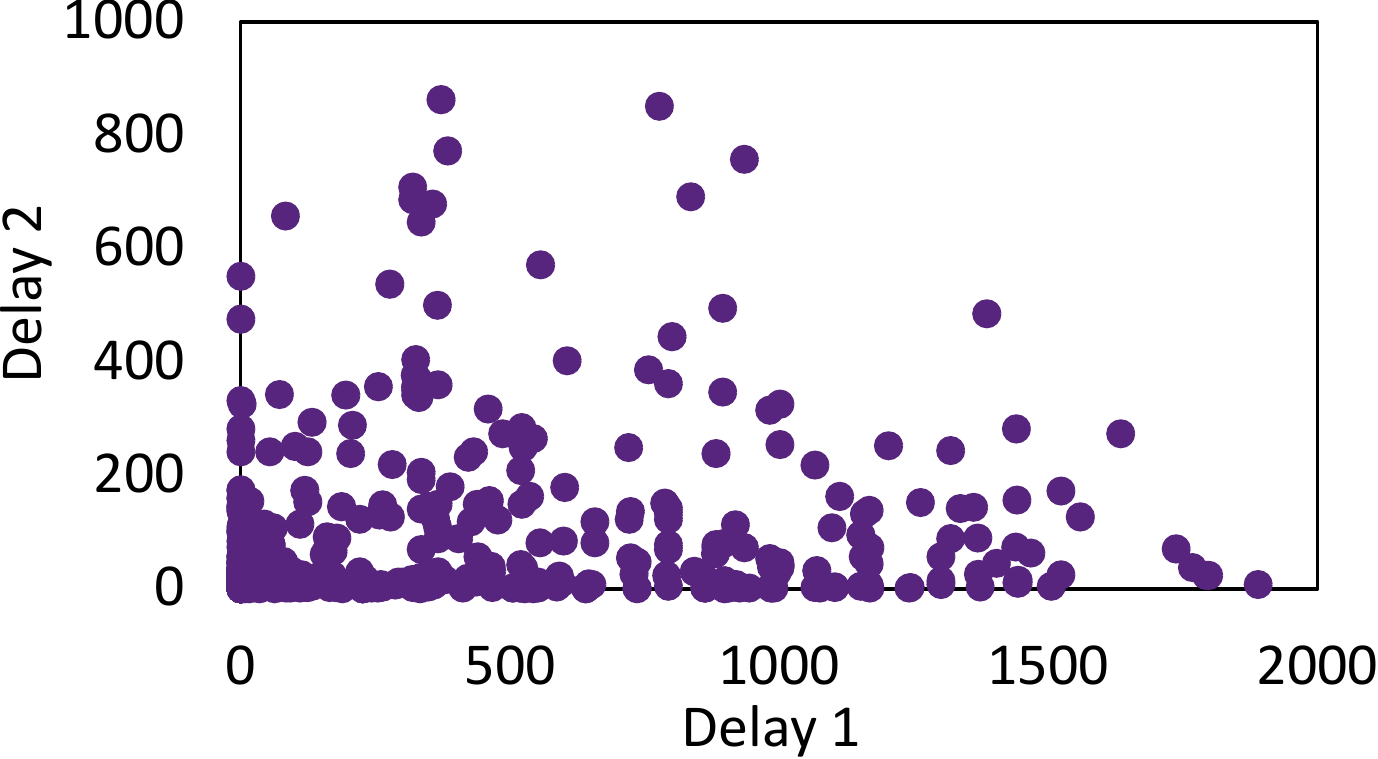}
        \caption{A scatter plot of $D_1$ and $D_2$.}
        \label{fig:delay_scatter}
    \end{subfigure}
    \vspace{-0.1in}
    \caption{Plots of $D_1$ and $D_2$}
    \vspace{-0.2in}
    \label{fig:delays}
\end{figure*}

\subsection{Bug Finding Delays}
\label{subsec:delays}

\begin{figure}
    \centering
    \includegraphics[width=0.9\columnwidth]{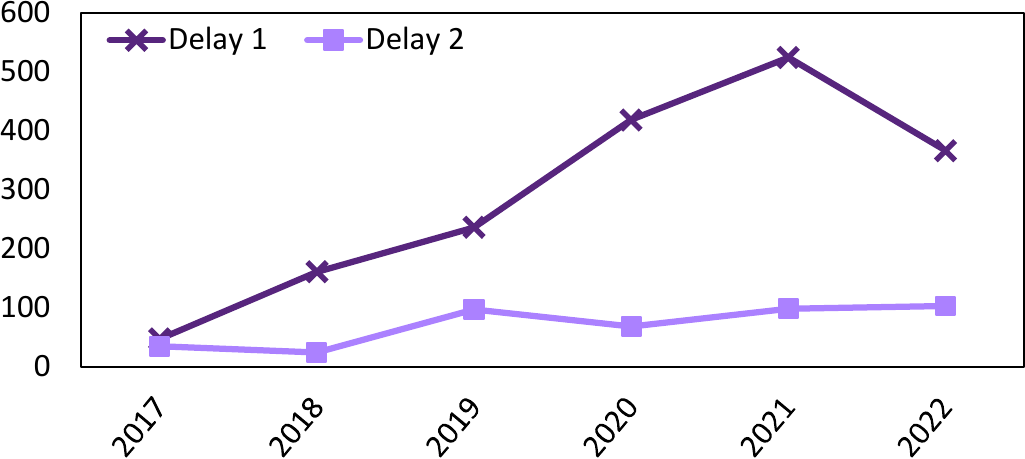}
    \caption{The average $D_1$ and $D_2$ over the past 6 years.}
    \label{fig:delay_by_year}
    \vspace{-0.2in}
\end{figure}

We now pivot to analyzing the bug finding delays $D_1$ and $D_2$.
Figures~\ref{fig:d1_hist} and~\ref{fig:d2_hist} show histograms of $D_1$ and $D_2$, which we observe are exponentially distributed.
Since for an exponential distribution all other metrics are derived from the mean, we will use the mean of the delays when describing them at large scales.
To this end, the average $D_1$ is $331.17$ days - nearly a year, while the average $D_2$ is only $73.94$ days.
In this section, we will disprove any correlation between $D_1$ and $D_2$, and then take a deeper dive into the trends of each delay.

We investigate whether $D_1$ and $D_2$ are independent.
To do so, we present Figure~\ref{fig:delay_scatter}, a scatter plot of $D_1$ and $D_2$, and obeserve that there is no obvious correlation.
We also perform regression analysis and find that no regression line, linear or curved, has an $r^2$ value greater than $0.126$.
For any $D_1$ we cannot predict the value of $D_2$.
We believe this is strong evidence that $D_1$ and $D_2$ are independent.
This has an important implication: it shows that separate efforts are needed to reduce each delay and provides a bound on how much each solution can reduce the delay.

\begin{tcolorbox}
    \textbf{Finding \nextfnum.}~~
    We find that $D_1$ and $D_2$ are independent components of the overall delay.
    This implies that separate efforts are needed to reduce $D_1$ and $D_2$.
\end{tcolorbox}

\subsubsection{Delay 1}
\label{subsubsec:delay1}
We first consider $D_1$ as shown in Figure~\ref{fig:delay_by_year}.
The increase in $D_1$ is expected as Syzbot begins fuzzing, noting that $D_1$ is measured from the time Syzbot began fuzzing for bugs older than it.
A general increase in $D_1$ means that more bugs are being found that were once hidden, and that those bugs are often older than the fuzzer.
We know that when Syzbot first began in 2017, Linux was already much older than it and had many bugs already.
So, Syzbot's increasing $D_1$ shows that it is improving.
This can be seen when comparing the slightly steeper jump of $D_1$ from 2019 to 2020 from Figure~\ref{fig:delay_by_year} to the increased number of reveals in the same year from Table~\ref{tab:reveal_by_year}.

\begin{table}
    \centering
    \scriptsize
    \caption{Number of bugs found in each year (columns) by the year they were introduced (rows). We note that 2005 is a hard barrier as this is the year Linux began using git.}
    \begin{tabular}{r|cccccc} \toprule
        Year & 2017 & 2018 & 2019 & 2020 & 2021 & 2022\\ \midrule
        2005 & 0 & 0 & 8 & 8 & 3 & 1\\
        2006 & 0 & 7 & 0 & 1 & 0 & 1\\
        2007 & 0 & 2 & 2 & 0 & 0 & 0\\
        2008 & 0 & 1 & 1 & 4 & 3 & 0\\
        2009 & 0 & 0 & 3 & 1 & 4 & 1\\
        2010 & 0 & 3 & 5 & 3 & 1 & 1\\
        2011 & 0 & 0 & 2 & 3 & 2 & 1\\
        2012 & 0 & 4 & 3 & 2 & 4 & 1\\
        2013 & 0 & 5 & 6 & 4 & 1 & 2\\
        2014 & 0 & 3 & 2 & 7 & 2 & 0\\
        2015 & 4 & 4 & 0 & 1 & 1 & 0\\
        2016 & 3 & 11 & 5 & 8 & 2 & 2\\
        2017 & 10 & 8 & 9 & 16 & 5 & 2\\
        2018 &  & 57 & 17 & 17 & 3 & 1\\
        2019 &  &  & 53 & 27 & 6 & 3\\
        2020 &  &  &  & 63 & 18 & 4\\
        2021 &  &  &  &  & 31 & 10\\
        2022 &  &  &  &  &  & 40\\
        \bottomrule
    \end{tabular}
    \label{tab:finding_by_guilty}
\end{table}

The upward trend of $D_1$ comes to an abrupt halt in 2022 when it takes a nose-dive.
We observe that a decrease in $D_1$ means the fuzzer is finding fewer previously hidden bugs, and that bugs are being revealed much earlier in their lifespans.
Indeed, Table~\ref{tab:finding_by_guilty} shows that bugs found in 2022 also largely originate from 2022.
This is in stark contrast to the previous years, which have a more even spread between bugs whose guilty commits lie in the same year and those that are much older.
One way or another, Syzbot is running out of old bugs to find.
Since these are only the bugs \textit{found} by Syzbot, it is difficult to make a definitive statement about what this particular decrease means.
It could be that Syzbot has found most of the bugs that pre-existed it and this is the payoff after years of hard work.
However, based on our observations of revealing factors (\S\ref{subsec:rf_spread}), this is quite unlikely.
It is more likely that Syzbot has yet to uncover a multitude of bugs somewhere in the kernel, bugs that are still hidden for various reasons (\eg needed kernel commits, other blocking bugs, missing syscall descriptions, \etc).
To this end, we observe that Syzbot has not achieved $100\%$ coverage of the Linux kernel, and syscall description development has slowed from its peak in 2020 (Figure~\ref{fig:template_churn}).
Because of this, we believe that Syzbot is approaching a kind of local ideal, where it has found most of the old bugs it is able to without major changes.
Since $D_1$ for Syzkaller commits, syscall description commits, and sanitizers is approaching $0$ as development slows, only $D_1$ for kernel commits and blocking bugs will remain.
Based on this and the distribution of revealing factors, we calculate that $D_1$ will settle at around $35\%$ of its peak value, or $187$ days.
This behavior is still important as it demonstrates one thing: \textit{Syzbot is capable of converging to an ideal state of fuzzing with the Linux kernel}.

\begin{tcolorbox}
    \textbf{Finding \nextfnum.}~~
    We find that based on the decreasing $D_1$, Syzbot is approaching a local ideal state of fuzzing, but that more improvement is still possible.
\end{tcolorbox}

\subsubsection{Delay 2}
\label{subsubsec:delay2}

$D_2$ is the delay for Syzbot to find a bug while the bug is revealed.
We reason that $D_2$ is largely related to the compute power given to Syzbot and the number of bugs in the kernel.
$D_2$ is what has been previously thought of as the bug finding delay or time-to-find.
If a fuzzer is given more resources (\ie compute power), it will fuzz the target with greater throughput, which in turn will find bugs faster.
In a sense, this defines the rate at which Syzbot can find bugs.
We will experimentally demonstrate this point in \S\ref{subsec:improve_d2}.
On the other hand, an increase in $D_2$ signifies that the fuzzer cannot keep up with the number of bugs being revealed. 
That is, if Syzbot finds bugs at the same rate, it takes longer on average to find 100 bugs rather than 50.
So, $D_2$ can be roughly thought of as Syzbot's fuzzing power, or how fast it can find bugs in a target.
We see one sizable increase in $D_2$ between the years 2018 and 2019 (Figure~\ref{fig:delay_by_year}).
During these years, Syzbot was unable to keep up with the number of bugs revealed to it.
After this, Syzbot's $D_2$ steadied at around $100$ days per bug.

\begin{tcolorbox}
    \textbf{Finding \nextfnum.}~~
    We find that $D_2$ is an approximate measure of Syzbot's fuzzing power, which (as we will show) is largely related to the compute power given to Syzbot.
\end{tcolorbox}

\subsection{Delays by Location}
\label{subsec:delay_by_loc}

To better understand bug finding delays, we look at how bug finding delays have changed for different locations in the kernel.
We will use the root directories of Linux (\ie drivers, net, kernel, \etc) to describe the possible locations where a bug could exist.
Further, we will specifically focus on the directories for which Syzbot found more than $50$ bugs, and which have substantial syscall development.
Those directories are drivers, the filesystem directory (FS), kernel, and network (net).

\begin{figure}
    \includegraphics[width=0.9\columnwidth]{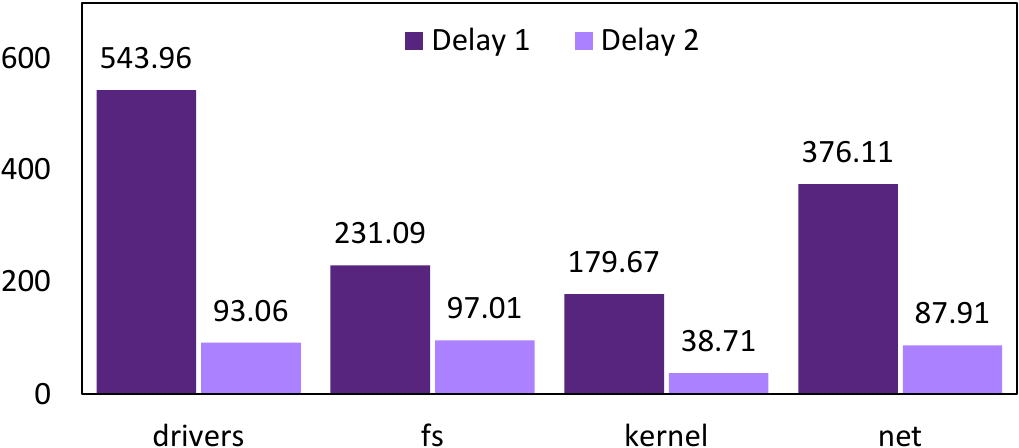}
    \caption{The average $D_1$ and $D_2$ in root directories of Linux.}
    \label{fig:delay_by_location}
    \vspace{-0.2in}
\end{figure}

Figure~\ref{fig:delay_by_location} shows the average delays of bugs split up by the directory they occur in.
We see a big difference in $D_1$ between the directories.
Consider the difference in $D_1$ between drivers, $543.96$ days, and net, $376.11$ days.
Both directories have seen substantial development regarding syscall descriptions, and both require some additional fuzzing support.
Drivers often require emulated or actual hardware which can be unique to each driver, and network interfaces often require assistance in setting up networks.
Despite these similarities, the drivers directory has proven to be much more difficult to reveal bugs in.
We note that the reason for the difference in $D_1$ is two-fold.
First is the aforementioned requirement of hardware or simulated hardware and the difficulty of writing syscall descriptions for them.
Net sees a similar issue, but has also received a fair amount of targeted development.
This however, only accounts for why a bug would take so long to be revealed by a Syzkaller or description commit.
The second reason is the size of the directory, which is positively correlated with $D_1$ (Table~\ref{tab:syscall_vs_loc}).
The larger the directory, the less likely that a kernel commit is going to influence the bug's findability.
So, bugs that are revealed by kernel commits or blocking bugs can be hidden for a long time as well.

\begin{table}
    \centering
    \small
    \caption{The percent of total lines of code in each root directory of the compiled kernel and the percent of syscall descriptions associated with each root directory.}
    \begin{tabular}{rcc} \toprule
        Directory & \% of Kernel LoC & \% of Descriptions\\ \midrule
        drivers & 41.6\% & 42.4\% \\
        fs & 17.1\% & 8.9\% \\
        kernel & 3.6\% & 9.7\% \\
        net & 27.0\% & 39.0\% \\
        \bottomrule
    \end{tabular}
    \vspace{-0.2in}
    \label{tab:syscall_vs_loc}
\end{table}

Let us compare the lines of code in each kernel directory with the number of syscall descriptions associated with each.
Table~\ref{tab:syscall_vs_loc} shows that despite drivers making up $1.5$ times the lines of code of the compiled kernel compared to net, there are only slightly more syscall descriptions of drivers than those of net.
Couple this with the long $D_1$ in drivers, and we see that drivers-based syscall descriptions need more development.
Indeed, syscall description development has dropped in recent years, and driver support has not seen major updates since 2019 (Figure~\ref{fig:template_churn}).
We believe that further development of driver related syscall descriptions will reveal many currently hidden bugs.
That said, we understand the difficulty in developing such descriptions, from the domain knowledge required to the hardware support that some drivers mandate.
This only serves to highlight the importance of research into tools which aim to automatically generate syscall descriptions for drivers~\cite{hao2023syzdescribe, corina2017difuze, shen2022drifuzz, bulekov2023no, sun2022ksg}.
We believe that this area of research is promising and will help fuzzer developers keep up with driver development.

\begin{tcolorbox}
    \textbf{Finding \nextfnum.}~~
    We find that the $D_1$ of a location is determined by both the size and complexity of that location.
\end{tcolorbox}

Figure~\ref{fig:delay_by_location} also reveals an interesting finding about $D_2$.
We see that drivers, FS, and net all share a similar $D_2$ (about $85$ to $100$ days), meaning Syzkaller does not struggle to find bugs in one location or another.
But we also see that the kernel directory has roughly half the $D_2$ of other directories at only $38.71$ days.
Kernel is a small directory and most test cases go through it during their execution, so it has a higher ``coverage frequency'' (\ie how often code is fuzzed).
This correlation shows that coverage frequency also plays a role in determining $D_2$.

\begin{tcolorbox}
    \textbf{Finding \nextfnum.}~~
    We find that the fuzzer's compute power is not distributed equally between different locations in the kernel.
    Some locations see a higher ``coverage frequency'' and hence a lower $D_2$.
\end{tcolorbox}

\section{Takeaways}
\label{sec:takeaways}

Based on our experimental results and understanding of continuous fuzzing, we provide some suggestions to improve the bug finding delay.

\subsection{Improving $D_1$}
\label{subsec:improve_d1}

In order to improve the $D_1$ of hard-to-induce reveals such as kernel commit reveals, we look to the core reason why these bugs are hidden.
There exists a buggy state in the kernel that does not trigger a crash, thus the fuzzer cannot find it.
So, we propose a solution that explicitly turns these buggy states into crashes.
Warning and bug assertions are already used in the kernel to identify specific buggy states.
We suggest making use of these and even automating their implementation in an effort to reveal bugs quicker.
One could imagine a tool which uses static analysis or an LLM to insert checks in high-risk areas.
The same strategy could be used to find classes of bugs that the fuzzer cannot yet find such as type confusions.
From our experience, Syzbot is already very good at triggering warning and bug assertions.
Recall a bug from \S\ref{subsubsec:kernel_commit} which was revealed and found due to an added warning.
Since the buggy state is a pipe direction mistake that quietly fails, it is hard to imagine the bug would ever be revealed except through a warning or bug assertion.
Furthermore, the bug was found only $2$ days after it was revealed.
We believe there are many more bugs like this that could be revealed through similar means.
In addition, bugs found this way may be easier to patch as the exact buggy state is already known.
We believe this line of work is promising for reducing the $D_1$ of otherwise hidden bugs.

\begin{tcolorbox}
    \textbf{Takeaway 1.}~~
    We suggest the use of automation, either static analysis or machine learning, to insert more warning and bug assertions and reveal more bugs.
\end{tcolorbox}

In order to improve Syzkaller, members of the community must identify areas that need work and use their domain knowledge to create syscall descriptions.
Here, we look for a method to better focus developers' time.
We surveyed $30$ recent CVEs and found that $13$ were unfindable by Syzbot.
$7$ of these bugs were unfindable because the proper syscall descriptions did not exist.
Consider CVE-2022-29156, a double free in the Infiniband RDMA Transport Server (RTRS) module.
The RTRS module is active in the kernel, and Syzkaller even has the proper syscalls to fuzz most of RDMA.
The correct syscall, \texttt{write\$RDMA\_USER\_CM\_CMD\_CONNECT}, is in Syzkaller's description set, but it does not have the proper flag, \texttt{RTRS\_MSG\_CON\_REQ}, to reach the RTRS module.
In our experience, it is a common trend that Syzbot cannot find CVEs due to an incomplete syscall description set.
We suggest taking advantage of this trend.
CVEs point to areas in the kernel where vulnerabilities exist, and Syzkaller needs more development.
This is an easy way to identify incomplete descriptions that are key to fuzzing vulnerable areas.

\begin{tcolorbox}
    \textbf{Takeaway 2.}~~
    We suggest using CVEs that are not found by Syzbot as an external trigger to develop syscall descriptions for vulnerable code.
\end{tcolorbox}

\subsection{Improving $D_2$}
\label{subsec:improve_d2}

The goal of Syzbot is to identify, track, and assist in removing bugs from Linux.
However, keeping up with kernel development is no small feat.
Linux makes major releases such as 5.14 or 5.15 every $63$ or $70$ days -- its release cycle.
We studied $90$ recent bugs and found that $76\%$ of bugs are patched within $70$ days of being found -- their time-to-fix.
To set an attainable goal, we focus only on $D_2$ as it is inpedendent of $D_1$.
Even with just $D_2$ and time-to-fix, bugs could exist in more than $3$ releases before they are patched.
We also note that due to the time-to-fix being as long as a release cycle, it is unreasonable to set a goal of removing all bugs from all releases.
Instead, we set a goal of $D_2 \leq 60$ days.
This would ensure that most bugs exist in at most $2$ releases.

\begin{figure}
    \centering
    \includegraphics[width=0.85\columnwidth]{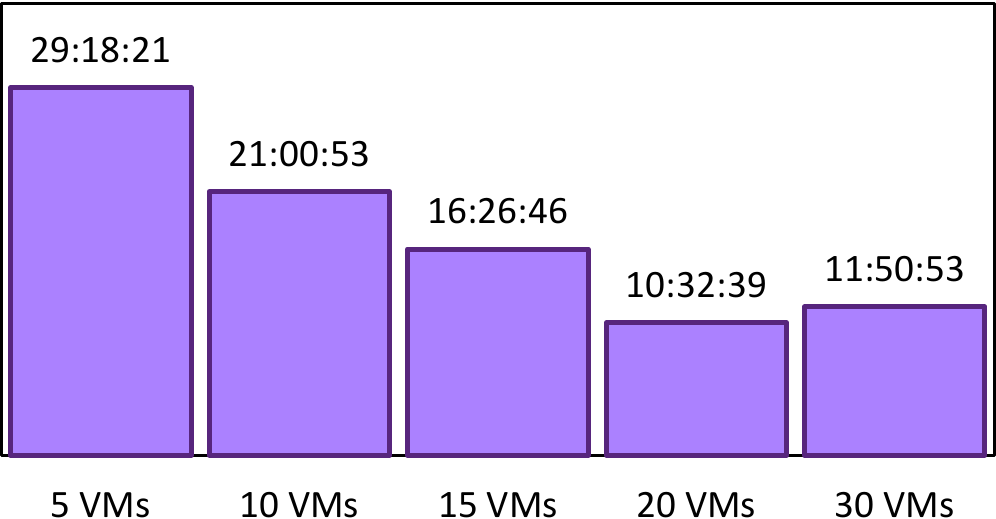}
    \caption{Average $D_2$ with varying numbers of VMs.}
    \label{fig:resource_est}
    \vspace{-0.2in}
\end{figure}

To reach this goal, we suggest adding more VMs to each manager under Syzbot.
A brief analysis of bugs found by Syzbot reveals that most bugs are found by a single manager.
This implies that while managers share test cases among each other, they work alone for the purposes of finding bugs.
To find out how many VMs should be added, we ran a small scale study of Syzkaller with varying number of VMs.
We note that each Syzbot manager uses $10$ VMs.
Figure~\ref{fig:resource_est} shows that there is a clear correlation between the number of VMs and $D_2$ up until $20$ VMs.
Coincidentally, this doubles the fuzzing power of Syzkaller while halving the $D_2$.
We estimate that doubling the number of VMs in each manager could cut $D_2$ down to just $50$ days, meeting our goal.

\begin{tcolorbox}
    \textbf{Takeaway 3.}~~
    To best fit the release cycle of Linux, we set a goal of $D_2 \leq 60$ days, and estimate that doubling the VMs used by Syzbot will meet this goal.
\end{tcolorbox}

\section{Threats to Validity}
\label{sec:threats}

In this section, we thoroughly present possible biases in our results from SyzRetrospector

\subsection{Bias 1: Unfound Bugs}
\label{subsec:unfound_bugs}

Despite our efforts, certain bugs cannot be completely retrospected and must be left out of our data set.
The majority of these bugs are ones that fail to reproduce in a timely manner.
Bugs that take longer than 30 minutes to reproduce could take several days to a week to complete retrospection, making a large scale study infeasible.
We observe that these bugs are largely races or other non-deterministic bugs with tight windows such that it is near impossible for Syzkaller to find them quickly and consistently.
Some of these race windows can be as small as 200 jiffies~\cite{unfoundbug9}.
Such bugs are not found at the finding commit, and thus cannot continue retrospection.
It follows that hard-to-find bugs are left out of this study.
However, we can generally reason that hard-to-find bugs have a longer $D_2$ in Syzbot.
As we demonstrated in our findings (\S\ref{subsubsec:delay2}), $D_1$ and $D_2$ are independent of each other, meaning only $D_2$ is underestimated.
This bias may affect findings 6 and 8, as they are drawn from $D_2$, and will affect our results presenting $D_2$, such that the true average $D_2$ is likely higher than our reported $D_2$.
For the subset of bugs that we did retrospect successfully, $D_2$ is correct.

The remainder of the unfound bugs suffered from build errors in either Syzkaller or Linux that remain unpatched.
However, we believe that the occurances of such errors are distributed randomly enough to not introduce a bias in our results.

\subsection{Bias 2: Repository Choice}
\label{subsec:repo_choice}

Our dataset is comprised of bugs that were found in mainline Linux (i.e., upstream).
The rationale behind this choice is two-fold.
First, it is the bugs that appear upstream that eventually make it into Linux releases and can be exploited in the wild.
Plenty of bugs are found during testing in auxiliary repositories and never make it to mainline.
So, we are less concerned with bugs that do not appear in mainline.
Second, auxiliary repositories such as \texttt{linux-next}, \texttt{net}, or \texttt{bpf}, are unreliable.
We found several cases where a bug's key commits no longer existed in the repository it was originally found in due to the repository being rebuilt.
If we cannot account for the bug's entire lifetime, we cannot perform accurate retrospection.
Therefore, we only take bugs which appear in the upstream repository.
This bias may affect findings 1, 2, and 3 as auxiliary repositories may have differing distributions.
Though we believe it to be unlikely, the distribution of revealing factors may be different in auxiliary repositories.

\begin{figure}
    \centering
    \vspace{0.1in}
    \includegraphics[width=\columnwidth]{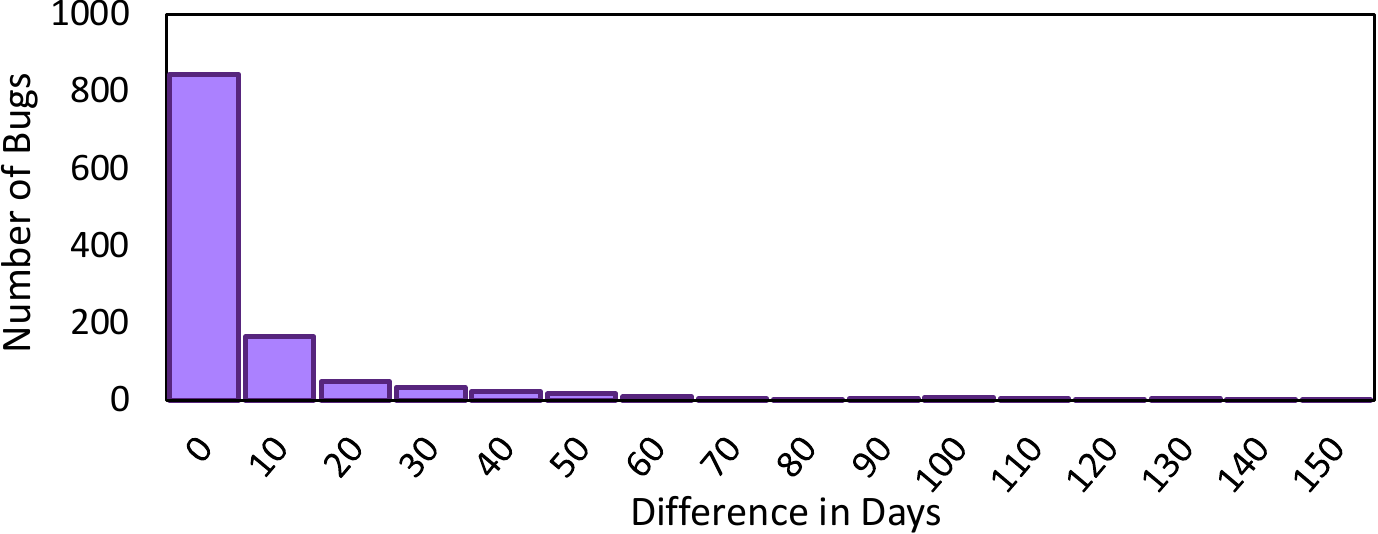}
    \caption{Finding date difference between upstream and auxiliary repositories. This graph covers the $1159$ potential bugs that SyzRetrospector attempted to retrospect.}
    \vspace{-0.2in}
    \label{fig:repo_date_diff}
\end{figure}

\subsection{Bias 3: Crash Instance Choice}
\label{subsec:crash_choice}

Our choice in repository also requires that the bug's finding commit be in upstream, which will determine the finding date of the bug.
This becomes an inaccuracy when a bug is found in another repository before it is found in mainline.
In this case the bug was found sooner than we give Syzbot credit for.
However, we analyzed this difference and see in Figure~\ref{fig:repo_date_diff} that most bugs found in auxiliary repositories are found within a few days in the upstream repository.
Well over half of the bugs are found on the same day, and over $84\%$ are found within $10$ days.
Couple this with the relatively long lifespan of bugs, and the difference in days becomes small.
As with Bias 1, this may affect findings 6 and 8, as well as our results on $D_2$, where it may reasonably vary $\pm 10$ days.

\section{Related Work}

\subsubsection{Previous Studies}
Despite the prominence of continuous fuzzers in project development, we have seen very few empirical studies of their performance.
Our survey of the subject area finds only two evaluations of continuous fuzzers, and only one related to Syzbot.
The study from Rouhonen et al.~\cite{ruohonen2019empirical} analyzes bugs found by Syzbot to understand their time-to-fix with regards to operating system and bug type.
Ding and Goues~\cite{ding2021empirical} perform a similar but more in-depth study of OSS-Fuzz~\cite{serebryany2017oss}.
The authors explore the idea of ``fuzz blocker'', very similar to our idea of blocking bugs (\S\ref{subsubsec:blocking_bugs}).
They also analyze the flakiness of bugs, whether the bug is patched, and whether it has a CVE.
Both of these studies rely on statistics and static analysis.
By contrast, our work with SyzRetrospector is the first large-scale, dynamic analysis of bug lifetimes with regard to Syzbot, or any continuous fuzzer.
We not only demonstrate that certain metrics like bug finding delay are inaccurate, but develop our own metrics to clearly analyze Syzbot's performance over $6$ years.

\subsubsection{Benchmark Tools}
Magma~\cite{hazimeh2020magma}, LAVA~\cite{dolan2016lava}, BugBench~\cite{lu2005bugbench}, FuzzBench~\cite{metzman2021fuzzbench}, and UNIFUZZ~\cite{li2021unifuzz}, are evaluation tools for a wide array of fuzzers based on how many and what types of bugs they can find out of a test suite.
While effective for conventional fuzzers, these benchmarks do not take into account the great lengths of time that continuous fuzzers take advantage of.
Because of this, current benchmarking tools are not a good fit for evaluating continuous fuzzers like Syzbot.

\subsubsection{Other Studies}
Additional studies have evaluated specific types of bugs or their security impacts.
SyzScope~\cite{zou2022syzscope} and KOOBE~\cite{chen2020koobe} analyze the impact of bugs, looking for high-risk consequences.
These works give a key reason to uncover and patch bugs quickly.
Other studies~\cite{weeratunge2010analyzing, bianchi2017reproducing, huang2013clap, machado2016production} use post processing on information like core dumps, stack traces, and thread execution logs to better understand bugs.
Still others~\cite{king2005debugging, srinivasan2004flashback, huang2010leap, lee2010respec} focus on record and replay to debug.
Execution Synthesis~\cite{zamfir2010execution} and Star~\cite{chen2014star} study concurrency bugs by reproducing them.
Our study is not concerned with a bug's impacts or its inner workings, rather we study its lifetime with relation to Syzbot.

\subsubsection{Strategies}
Bisection is well known throughout the community and closely related to our strategies in SyzRetrospector.
It has been implemented in git~\cite{git_bisect}, improved upon in some aspects~\cite{saha2017selective}, and is already used in Syzbot to assist developers~\cite{syzbot_bisection}.
Our tool builds on bisection to include Syzkaller and syscall description commits to determine when a bug was revealed, not its root cause.

SyzDescribe~\cite{hao2023syzdescribe}, DriFuzz~\cite{shen2022drifuzz}, Syzgen~\cite{chen2021syzgen}, DIFUZE~\cite{corina2017difuze}, and more~\cite{liu2020fans, bulekov2023no, sun2022ksg}, provide solutions for fuzzing kernel interfaces without syscall descriptions by automatically generating their own descriptions.
Each of these works has the potential to be invaluable for fuzzing areas of code that lack syscall descriptions.
Though Difuze was integrated into Syzkaller for testing on Android, Syzkaller still depends on manual description generation for Linux.

Bowknots~\cite{talebi2021undo} and Talos~\cite{huang2016talos} both provide workarounds for found bugs that are not patched yet.
These strategies represent possible solutions for blocking bugs until they are patched out, though they are not in use with Syzbot.

\subsubsection{Other Fuzzers}
Much previous work has been done in fuzzing in recent years.
Directed fuzzers~\cite{bohme2017directed, osterlund2020parmesan, zhu2021regression, AFL, fioraldi2020afl++, lemieux2018fairfuzz} efficiently target and find bugs, usually based on distance to target.
Hybrid fuzzers~\cite{yun2018qsym, stephens2016driller, chen2018angora, yue2020ecofuzz, she2019neuzz, wang2021syzvegas, godefroid2005dart} use symbolic execution and other heuristics to improve time-to-find.
Other fuzzers focus on concurrency bugs~\cite{jeong2019razzer, xu2020krace, fonseca2014ski}, often controlling how threads are interleaved.
Our contribution, focused fuzzing, is different in that we remove some syscall descriptions from Syzkaller without changing the fuzzing algorithm.
Furthermore, focused fuzzing is not meant to compete with these fuzzers.
Indeed directed fuzzers are faster and more reliable than focused fuzzing, but using them would nullify our ability to retrospect Syzbot.

\section{Conclusion}

In a short pilot study, we found that commonly used evaluation metrics are inaccurate since bugs often spend the majority of their lives hidden.
In response to this, we undertook a large-scale study to provide a better understanding and quantification of Syzbot's bug-finding performance and improvements.
Our tool, SyzRetrospector, takes a different approach to evaluating Syzbot by dynamically searching for the earliest that Syzbot was capable of finding a bug, and why that bug was revealed.
We used SyzRetrospector to analyze $559$ bugs and found that bugs are hidden for an average of $331.17$ days before they are revealed.
We present findings on the behaviors of revealing factors, how some bugs are harder to reveal than others, the trends in delays over the past $6$ years, and how bug location relates to delays.
Finally, we provide takeaways for improving Syzbot's delays based on our findings and experience.

\section*{Acknowledgments}

The work was supported in part by NSF Awards \#1953932, \#2247880, and \#2155213 as well as NSA Award \#H98230-22-1-0308.

\balance
\bibliographystyle{IEEEtranS}
\bibliography{ardalan}

\begin{thebibliography}{10}
\providecommand{\url}[1]{#1}
\csname url@samestyle\endcsname
\providecommand{\newblock}{\relax}
\providecommand{\bibinfo}[2]{#2}
\providecommand{\BIBentrySTDinterwordspacing}{\spaceskip=0pt\relax}
\providecommand{\BIBentryALTinterwordstretchfactor}{4}
\providecommand{\BIBentryALTinterwordspacing}{\spaceskip=\fontdimen2\font plus
\BIBentryALTinterwordstretchfactor\fontdimen3\font minus
  \fontdimen4\font\relax}
\providecommand{\BIBforeignlanguage}[2]{{%
\expandafter\ifx\csname l@#1\endcsname\relax
\typeout{** WARNING: IEEEtranS.bst: No hyphenation pattern has been}%
\typeout{** loaded for the language `#1'. Using the pattern for}%
\typeout{** the default language instead.}%
\else
\language=\csname l@#1\endcsname
\fi
#2}}
\providecommand{\BIBdecl}{\relax}
\BIBdecl

\bibitem{syz-env}
``{How to contribute to syzkaller},''
  \url{https://github.com/google/syzkaller/blob/master/docs/contributing.md},
  2018.

\bibitem{git_bisect}
``{Git Bisect},'' \url{https://git-scm.com/docs/git-bisect}, 2020.

\bibitem{AFL}
``{american fuzzy lop},'' \url{https://github.com/google/AFL}, 2021.

\bibitem{pilotbug4}
``{memory leak in kobject\_set\_name\_vargs (4)},''
  \url{https://syzkaller.appspot.com/bug?id=89c3ddb9936d3552995130298f1d2633ab9d3541},
  2021.

\bibitem{pilotbug9}
``{WARNING in exception\_type},''
  \url{https://syzkaller.appspot.com/bug?id=dccafc201251e8dfa52f17986d33f7ecbd6747fc},
  2021.

\bibitem{pilotbug10}
``{WARNING in futex\_requeue},''
  \url{https://syzkaller.appspot.com/bug?id=03f29b6252786a6f17661d727c03c83a7f70c86e},
  2021.

\bibitem{pilotbug1}
``{WARNING in rtl28xxu\_ctrl\_msg/usb\_submit\_urb},''
  \url{https://syzkaller.appspot.com/bug?id=e98d2e8aa7283d11aa8e0b718d8afa1a058e6ae0},
  2021.

\bibitem{pilotbug12}
``{KASAN: slab-out-of-bounds Read in packet\_recvmsg},''
  \url{https://syzkaller.appspot.com/bug?id=7c7245f9088053e9e49b97a341dee26c9ed40a2c},
  2022.

\bibitem{unexpectbug111}
``{KASAN: slab-out-of-bounds Read in thrustmaster\_probe},''
  \url{https://syzkaller.appspot.com/bug?id=e1c3525a4f4e2e4b6c1f73611ceaf61ef462700c},
  2022.

\bibitem{kmsanbug9}
``{KMSAN: uninit-value in number (4)},''
  \url{https://syzkaller.appspot.com/bug?id=59384a424f10c52eba52e098087f428e3a8b1915},
  2022.

\bibitem{syzbot}
``{syzbot},'' \url{https://syzkaller.appspot.com/upstream}, 2022.

\bibitem{syzbot_syzlang}
``{Syzkaller: Syzcall Description Language},''
  \url{https://github.com/google/syzkaller/blob/master/docs/syscall\_descriptions\_syntax.md},
  2022.

\bibitem{casestudybug5}
``{WARNING in dlfb\_submit\_urb/usb\_submit\_urb},''
  \url{https://syzkaller.appspot.com/bug?id=9c2df342be9d102da75f9532e168a95b9c379ae4},
  2022.

\bibitem{kmemleak2023}
``{Kernel Memory Leak Detector},''
  \url{https://docs.kernel.org/dev-tools/kmemleak.html}, 2023.

\bibitem{kasan2023}
``{The Kernel Address Sanitizer (KASAN)},''
  \url{https://docs.kernel.org/dev-tools/kasan.html}, 2023.

\bibitem{kcsan2023}
``{The Kernel Concurrency Sanitizer (KCSAN)},''
  \url{https://docs.kernel.org/dev-tools/kcsan.html}, 2023.

\bibitem{kmsan2023}
``{The Kernel Memory Sanitizer (KMSAN)},''
  \url{https://docs.kernel.org/dev-tools/kmsan.html}, 2023.

\bibitem{ubsan2023}
``{The Undefined Behavior Sanitizer - UBSAN},''
  \url{https://docs.kernel.org/dev-tools/ubsan.html}, 2023.

\bibitem{bianchi2017reproducing}
F.~A. Bianchi, M.~Pezz{\`e}, and V.~Terragni, ``Reproducing concurrency
  failures from crash stacks,'' in \emph{Proceedings of the 2017 11th Joint
  Meeting on Foundations of Software Engineering}, 2017, pp. 705--716.

\bibitem{bohme2017directed}
M.~B{\"o}hme, V.-T. Pham, M.-D. Nguyen, and A.~Roychoudhury, ``Directed greybox
  fuzzing,'' in \emph{Proceedings of the 2017 ACM SIGSAC Conference on Computer
  and Communications Security}, 2017, pp. 2329--2344.

\bibitem{bulekov2023no}
A.~Bulekov, B.~Das, S.~Hajnoczi, and M.~Egele, ``No grammar, no problem:
  Towards fuzzing the linux kernel without system-call descriptions.'' in
  \emph{NDSS}, 2023.

\bibitem{chen2014star}
N.~Chen and S.~Kim, ``Star: Stack trace based automatic crash reproduction via
  symbolic execution,'' \emph{IEEE transactions on software engineering},
  vol.~41, no.~2, pp. 198--220, 2014.

\bibitem{chen2018angora}
P.~Chen and H.~Chen, ``Angora: Efficient fuzzing by principled search,'' in
  \emph{2018 IEEE Symposium on Security and Privacy (SP)}.\hskip 1em plus 0.5em
  minus 0.4em\relax IEEE, 2018, pp. 711--725.

\bibitem{chen2021syzgen}
W.~Chen, Y.~Wang, Z.~Zhang, and Z.~Qian, ``Syzgen: Automated generation of
  syscall specification of closed-source macos drivers,'' in \emph{ACM CCS},
  2021.

\bibitem{chen2020koobe}
W.~Chen, X.~Zou, G.~Li, and Z.~Qian, ``Koobe: Towards facilitating exploit
  generation of kernel out-of-bounds write vulnerabilities,'' in
  \emph{Proceedings of the 29th USENIX Conference on Security Symposium}, 2020,
  pp. 1093--1110.

\bibitem{corina2017difuze}
J.~Corina, A.~Machiry, C.~Salls, Y.~Shoshitaishvili, S.~Hao, C.~Kruegel, and
  G.~Vigna, ``Difuze: Interface aware fuzzing for kernel drivers,'' in
  \emph{Proceedings of the 2017 ACM SIGSAC Conference on Computer and
  Communications Security}, 2017, pp. 2123--2138.

\bibitem{ding2021empirical}
Z.~Y. Ding and C.~Le~Goues, ``An empirical study of oss-fuzz bugs,'' in
  \emph{2021 IEEE/ACM 18th International Conference on Mining Software
  Repositories (MSR)}.\hskip 1em plus 0.5em minus 0.4em\relax IEEE, 2021, pp.
  131--142.

\bibitem{dolan2016lava}
B.~Dolan-Gavitt, P.~Hulin, E.~Kirda, T.~Leek, A.~Mambretti, W.~Robertson,
  F.~Ulrich, and R.~Whelan, ``Lava: Large-scale automated vulnerability
  addition,'' in \emph{2016 IEEE symposium on security and privacy (SP)}.\hskip
  1em plus 0.5em minus 0.4em\relax IEEE, 2016, pp. 110--121.

\bibitem{fioraldi2020afl++}
A.~Fioraldi, D.~Maier, H.~Ei{\ss}feldt, and M.~Heuse, ``Afl++ combining
  incremental steps of fuzzing research,'' in \emph{Proceedings of the 14th
  USENIX Conference on Offensive Technologies}, 2020, pp. 10--10.

\bibitem{fonseca2014ski}
P.~Fonseca, R.~Rodrigues, and B.~B. Brandenburg, ``$\{$SKI$\}$: Exposing kernel
  concurrency bugs through systematic schedule exploration,'' in \emph{11th
  $\{$USENIX$\}$ Symposium on Operating Systems Design and Implementation
  ($\{$OSDI$\}$ 14)}, 2014, pp. 415--431.

\bibitem{godefroid2005dart}
P.~Godefroid, N.~Klarlund, and K.~Sen, ``Dart: Directed automated random
  testing,'' in \emph{Proceedings of the 2005 ACM SIGPLAN conference on
  Programming language design and implementation}, 2005, pp. 213--223.

\bibitem{hao2023syzdescribe}
Y.~Hao, G.~Li, X.~Zou, W.~Chen, S.~Zhu, Z.~Qian, and A.~A. Sani, ``Syzdescribe:
  Principled, automated, static generation of syscall descriptions for kernel
  drivers,'' in \emph{2023 IEEE Symposium on Security and Privacy (SP)}.\hskip
  1em plus 0.5em minus 0.4em\relax IEEE Computer Society, 2023, pp. 3262--3278.

\bibitem{hazimeh2020magma}
A.~Hazimeh, A.~Herrera, and M.~Payer, ``Magma: A ground-truth fuzzing
  benchmark,'' \emph{Proceedings of the ACM on Measurement and Analysis of
  Computing Systems}, vol.~4, no.~3, pp. 1--29, 2020.

\bibitem{kernelpatch1}
{H.J. Lu}, ``{x86/build/64: Force the linker to use 2MB page size},''
  \url{https://git.kernel.org/pub/scm/linux/kernel/git/torvalds/linux.git/commit/?id=e3d03598e8ae7d195af5d3d049596dec336f569f},
  2018.

\bibitem{huang2010leap}
J.~Huang, P.~Liu, and C.~Zhang, ``Leap: Lightweight deterministic
  multi-processor replay of concurrent java programs,'' in \emph{Proceedings of
  the eighteenth ACM SIGSOFT international symposium on Foundations of software
  engineering}, 2010, pp. 207--216.

\bibitem{huang2013clap}
J.~Huang, C.~Zhang, and J.~Dolby, ``Clap: Recording local executions to
  reproduce concurrency failures,'' \emph{Acm Sigplan Notices}, vol.~48, no.~6,
  pp. 141--152, 2013.

\bibitem{huang2016talos}
Z.~Huang, M.~DAngelo, D.~Miyani, and D.~Lie, ``Talos: Neutralizing
  vulnerabilities with security workarounds for rapid response,'' in \emph{2016
  IEEE Symposium on Security and Privacy (SP)}.\hskip 1em plus 0.5em minus
  0.4em\relax IEEE, 2016, pp. 618--635.

\bibitem{jeong2019razzer}
D.~R. Jeong, K.~Kim, B.~Shivakumar, B.~Lee, and I.~Shin, ``Razzer: Finding
  kernel race bugs through fuzzing,'' in \emph{2019 IEEE Symposium on Security
  and Privacy (SP)}.\hskip 1em plus 0.5em minus 0.4em\relax IEEE, 2019, pp.
  754--768.

\bibitem{king2005debugging}
S.~T. King, G.~W. Dunlap, and P.~M. Chen, ``Debugging operating systems with
  time-traveling virtual machines,'' in \emph{Proceedings of the 2005 USENIX
  Technical Conference}, 2005, pp. 1--15.

\bibitem{lee2010respec}
D.~Lee, B.~Wester, K.~Veeraraghavan, S.~Narayanasamy, P.~M. Chen, and J.~Flinn,
  ``Respec: efficient online multiprocessor replayvia speculation and external
  determinism,'' \emph{ACM Sigplan Notices}, vol.~45, no.~3, pp. 77--90, 2010.

\bibitem{lemieux2018fairfuzz}
C.~Lemieux and K.~Sen, ``Fairfuzz: A targeted mutation strategy for increasing
  greybox fuzz testing coverage,'' in \emph{Proceedings of the 33rd ACM/IEEE
  International Conference on Automated Software Engineering}, 2018, pp.
  475--485.

\bibitem{li2021unifuzz}
Y.~Li, S.~Ji, Y.~Chen, S.~Liang, W.-H. Lee, Y.~Chen, C.~Lyu, C.~Wu, R.~Beyah,
  P.~Cheng \emph{et~al.}, ``Unifuzz: A holistic and pragmatic metrics-driven
  platform for evaluating fuzzers.'' in \emph{USENIX Security Symposium}, 2021,
  pp. 2777--2794.

\bibitem{liu2020fans}
B.~Liu, C.~Zhang, G.~Gong, Y.~Zeng, H.~Ruan, and J.~Zhuge, ``Fans: Fuzzing
  android native system services via automated interface analysis.'' in
  \emph{USENIX Security Symposium}, 2020, pp. 307--323.

\bibitem{lu2005bugbench}
S.~Lu, Z.~Li, F.~Qin, L.~Tan, P.~Zhou, and Y.~Zhou, ``Bugbench: Benchmarks for
  evaluating bug detection tools,'' in \emph{Workshop on the evaluation of
  software defect detection tools}, vol.~5.\hskip 1em plus 0.5em minus
  0.4em\relax Chicago, Illinois, 2005.

\bibitem{machado2016production}
N.~Machado, B.~Lucia, and L.~Rodrigues, ``Production-guided concurrency
  debugging,'' in \emph{Proceedings of the 21st ACM SIGPLAN Symposium on
  Principles and Practice of Parallel Programming}, 2016, pp. 1--12.

\bibitem{metzman2021fuzzbench}
J.~Metzman, L.~Szekeres, L.~Simon, R.~Sprabery, and A.~Arya, ``Fuzzbench: an
  open fuzzer benchmarking platform and service,'' in \emph{Proceedings of the
  29th ACM joint meeting on European software engineering conference and
  symposium on the foundations of software engineering}, 2021, pp. 1393--1403.

\bibitem{osterlund2020parmesan}
S.~{\"O}sterlund, K.~Razavi, H.~Bos, and C.~Giuffrida, ``Parmesan:
  Sanitizer-guided greybox fuzzing,'' in \emph{Proceedings of the 29th USENIX
  Conference on Security Symposium}, 2020, pp. 2289--2306.

\bibitem{unfoundbug9}
{Peilin Ye}, ``{vsock: Fix memory leak in vsock\_connect()},''
  \url{https://git.kernel.org/pub/scm/linux/kernel/git/torvalds/linux.git/commit/?id=7e97cfed9929eaabc41829c395eb0d1350fccb9d},
  2022.

\bibitem{unexpectreveal111}
{Qu Wenruo}, ``{Revert "btrfs: compression: don't try to compress if we don't
  have enough pages"},''
  \url{https://git.kernel.org/pub/scm/linux/kernel/git/torvalds/linux.git/commit/?id=4e9655763b82a91e4c341835bb504a2b1590f984},
  2021.

\bibitem{ruohonen2019empirical}
J.~Ruohonen and K.~Rindell, ``Empirical notes on the interaction between
  continuous kernel fuzzing and development,'' in \emph{2019 IEEE International
  Symposium on Software Reliability Engineering Workshops (ISSREW)}.\hskip 1em
  plus 0.5em minus 0.4em\relax IEEE, 2019, pp. 276--281.

\bibitem{saha2017selective}
R.~Saha and M.~Gligoric, ``Selective bisection debugging,'' in
  \emph{Fundamental Approaches to Software Engineering: 20th International
  Conference, FASE 2017, Held as Part of the European Joint Conferences on
  Theory and Practice of Software, ETAPS 2017, Uppsala, Sweden, April 22-29,
  2017, Proceedings 20}.\hskip 1em plus 0.5em minus 0.4em\relax Springer, 2017,
  pp. 60--77.

\bibitem{serebryany2017oss}
K.~Serebryany, ``Oss-fuzz-google's continuous fuzzing service for open source
  software,'' in \emph{USENIX Security symposium}.\hskip 1em plus 0.5em minus
  0.4em\relax USENIX Association, 2017.

\bibitem{she2019neuzz}
D.~She, K.~Pei, D.~Epstein, J.~Yang, B.~Ray, and S.~Jana, ``Neuzz: Efficient
  fuzzing with neural program smoothing,'' in \emph{2019 IEEE Symposium on
  Security and Privacy (SP)}.\hskip 1em plus 0.5em minus 0.4em\relax IEEE,
  2019, pp. 803--817.

\bibitem{shen2022drifuzz}
Z.~Shen, R.~Roongta, and B.~Dolan-Gavitt, ``Drifuzz: Harvesting bugs in device
  drivers from golden seeds,'' in \emph{31st USENIX Security Symposium (USENIX
  Security 22)}, 2022, pp. 1275--1290.

\bibitem{srinivasan2004flashback}
S.~M. Srinivasan, S.~Kandula, C.~R. Andrews, Y.~Zhou \emph{et~al.},
  ``Flashback: A lightweight extension for rollback and deterministic replay
  for software debugging,'' in \emph{USENIX Annual Technical Conference,
  General Track}.\hskip 1em plus 0.5em minus 0.4em\relax Boston, MA, USA, 2004,
  pp. 29--44.

\bibitem{stephens2016driller}
N.~Stephens, J.~Grosen, C.~Salls, A.~Dutcher, R.~Wang, J.~Corbetta,
  Y.~Shoshitaishvili, C.~Kruegel, and G.~Vigna, ``Driller: Augmenting fuzzing
  through selective symbolic execution.'' in \emph{NDSS}, vol.~16, no. 2016,
  2016, pp. 1--16.

\bibitem{sun2022ksg}
H.~Sun, Y.~Shen, J.~Liu, Y.~Xu, and Y.~Jiang, ``$\{$KSG$\}$: Augmenting kernel
  fuzzing with system call specification generation,'' in \emph{2022 USENIX
  Annual Technical Conference (USENIX ATC 22)}, 2022, pp. 351--366.

\bibitem{talebi2021undo}
S.~M.~S. Talebi, Z.~Yao, A.~A. Sani, Z.~Qian, and D.~Austin, ``Undo workarounds
  for kernel bugs,'' in \emph{USENIX Security Symposium}, 2021.

\bibitem{syzbot_bisection}
D.~Vyukov, ``{syzbot Bisection},''
  \url{https://github.com/google/syzkaller/blob/master/docs/syzbot.md\#bisection},
  2018.

\bibitem{syzkallercommit1}
{Vyukov, D.}, ``{sys/linux: add syz\_mount\_image for 20 more file systems},''
  \url{https://github.com/google/syzkaller/commit/8394d04bf63f992da0a762fd987693521c2e2507},
  2020.

\bibitem{wang2021syzvegas}
D.~Wang, Z.~Zhang, H.~Zhang, Z.~Qian, S.~V. Krishnamurthy, and N.~B.
  Abu-Ghazaleh, ``Syzvegas: Beating kernel fuzzing odds with reinforcement
  learning.'' in \emph{USENIX Security Symposium}, 2021, pp. 2741--2758.

\bibitem{weeratunge2010analyzing}
D.~Weeratunge, X.~Zhang, and S.~Jagannathan, ``Analyzing multicore dumps to
  facilitate concurrency bug reproduction,'' in \emph{Proceedings of the
  fifteenth International Conference on Architectural support for programming
  languages and operating systems}, 2010, pp. 155--166.

\bibitem{xu2020krace}
M.~Xu, S.~Kashyap, H.~Zhao, and T.~Kim, ``Krace: Data race fuzzing for kernel
  file systems,'' in \emph{2020 IEEE Symposium on Security and Privacy
  (SP)}.\hskip 1em plus 0.5em minus 0.4em\relax IEEE, 2020, pp. 1643--1660.

\bibitem{yue2020ecofuzz}
T.~Yue, P.~Wang, Y.~Tang, E.~Wang, B.~Yu, K.~Lu, and X.~Zhou, ``Ecofuzz:
  Adaptive energy-saving greybox fuzzing as a variant of the adversarial
  multi-armed bandit,'' in \emph{Proceedings of the 29th USENIX Conference on
  Security Symposium}, 2020, pp. 2307--2324.

\bibitem{yun2018qsym}
I.~Yun, S.~Lee, M.~Xu, Y.~Jang, and T.~Kim, ``$\{$QSYM$\}$: A practical
  concolic execution engine tailored for hybrid fuzzing,'' in \emph{27th
  $\{$USENIX$\}$ Security Symposium ($\{$USENIX$\}$ Security 18)}, 2018, pp.
  745--761.

\bibitem{zamfir2010execution}
C.~Zamfir and G.~Candea, ``Execution synthesis: a technique for automated
  software debugging,'' in \emph{Proceedings of the 5th European conference on
  Computer systems}, 2010, pp. 321--334.

\bibitem{zhu2021regression}
X.~Zhu and M.~B{\"o}hme, ``Regression greybox fuzzing,'' in \emph{Proceedings
  of the 2021 ACM SIGSAC Conference on Computer and Communications Security},
  2021, pp. 2169--2182.

\bibitem{zou2022syzscope}
X.~Zou, G.~Li, W.~Chen, H.~Zhang, and Z.~Qian, ``$\{$SyzScope$\}$: Revealing
  $\{$High-Risk$\}$ security impacts of $\{$Fuzzer-Exposed$\}$ bugs in linux
  kernel,'' in \emph{31st USENIX Security Symposium (USENIX Security 22)},
  2022, pp. 3201--3217.

\end{thebibliography}

\end{document}